\documentclass[11pt,a4paper,english]{article}
\usepackage[T1]{fontenc}
\usepackage[latin1]{inputenc}
\usepackage{subfigure}
\usepackage{amsmath}
\usepackage{graphicx}

\makeatletter
\usepackage{a4wide}
\usepackage{psfrag}
\makeatletter
\renewcommand{\theequation}{\hbox{\normalsize\arabic{section}.\arabic{equation}}}
\@addtoreset{equation}{section}
\renewcommand{\thefigure}{\hbox{\normalsize\arabic{section}.\arabic{figure}}}
\@addtoreset{figure}{section}
\renewcommand{\thetable}{\hbox{\normalsize\arabic{section}.\arabic{table}}}
\@addtoreset{table}{section}
\makeatother 

\usepackage{babel}
\makeatother
\begin{document}

\title{\begin{flushright}{\normalsize ITP-Budapest Report No. 635}\end{flushright}\vspace{1cm}Form
factors in finite volume II:\\
disconnected terms and finite temperature correlators}

\author{B. Pozsgay$^{1}$%
\thanks{E-mail: pozsi@bolyai.elte.hu%
} ~and G. Takács$^{2}$%
\thanks{E-mail: takacs@elte.hu%
}\\
\\
$^{1}$\emph{Institute for Theoretical Physics }\\
\emph{Eötvös University, Budapest}\\
\emph{}\\
$^{2}$\emph{HAS Research Group for Theoretical Physics}\\
\emph{H-1117 Budapest, Pázmány Péter sétány 1/A}}

\date{25th June 2007}

\maketitle
\begin{abstract}
Continuing the investigation started in a previous work, we consider
form factors of integrable quantum field theories in finite volume,
extending our investigation to matrix elements with disconnected pieces.
Numerical verification of our results is provided by truncated conformal
space approach. Such matrix elements are important in computing finite
temperature correlation functions, and we give a new method for generating
a low temperature expansion, which we test for the one-point function
up to third order. 
\end{abstract}

\section{Introduction}

The matrix elements of local operators, the so-called form factors
are central objects in quantum field theory. In two-dimensional integrable
quantum field theory, the $S$ matrix can be obtained exactly in the
framework of factorized scattering (see \cite{ZZ79,Sbstr} for reviews).
Using the scattering amplitudes as input, it is possible to obtain
a set of axioms \cite{karowski} which provides the basis for the
form factor bootstrap (see \cite{Smirnov} for a review). 

Although in the bootstrap approach the connection with the Lagrangian
formulation of quantum field theory is rather indirect, it is thought
that the general solution of the form factor axioms determines the
complete local operator algebra of the theory \cite{cardy_mussardo},
which was confirmed in many cases by explicit comparison of the space
of solutions to the spectrum of local operators \cite{koubek_mussardo,koubek1,koubek2,smirnovcounting}.
Another important piece of information comes from correlation functions:
using form factors, a spectral representation for the correlation
functions can be built which provides a large distance expansion \cite{isingspincorr,Z1},
while the Lagrangian or perturbed conformal field theory formulation
allows one to obtain a short-distance expansion, which can then be
compared provided there is an overlap between their regimes of validity
\cite{Z1}. Other evidence for the correspondence between the field
theory and the solutions of the form factor bootstrap results from
evaluating sum rules like Zamolodchikov's $c$-theorem \cite{c-th,c-thspectral}
or the $\Delta$-theorem \cite{delta-th}, both of which can be used
to express conformal data as spectral sums in terms of form-factors.
Direct comparisons with multi-particle matrix elements are not so
readily available, except for perturbative or $1/N$ calculations
in some simple cases \cite{karowski}. One of our aims is to provide
non-perturbative evaluation of form factors from the Hamiltonian formulation,
which then allows for a direct comparison with solutions of the form
factor axioms.

Based on what we learned from our previous investigation of decay
rates in finite volume \cite{takacspozsgay}, in our previous paper
\cite{fftcsa} we determined form factors using a formulation of the
field theory in finite volume. We used the truncated conformal space
approach (TCSA) developed by Yurov and Al.B. Zamolodchikov \cite{yurov_zamolodchikov}
as a basis for numerical comparison to non-perturbative Hamiltonian
formulation of quantum field theory, and also its fermionic version
in the case of the Ising model \cite{tfcsa}. We were able to give
an extensive and direct numerical comparison between bootstrap results
for form factors and matrix elements evaluated non-perturbatively.
One of the advantages is that we can compare matrix elements directly,
without using any proxy (such as a two-point function or a sum rule);
the other is the very high precision of the comparison and also that
it is possible to test form factors of many particles which have never
been tested using spectral sums. Our approach, in contrast, makes
it possible to test entire one-dimensional sections of the form factor
functions using the volume as a parameter, and the number of available
sections only depends on our ability to identify multi-particle states
in finite volume. Part of the motivation of this work is to complete
the non-perturbative evaluation of form factors by extending our results
to matrix elements with disconnected pieces. 

Another motivation is provided by the fact that such matrix elements
are relevant for the calculation of finite temperature correlators.
Finite temperature correlation functions have attracted quite a lot
of interest recently \cite{sachdev,leclairmussardo,saleurfiniteT,lukyanovfiniteT,delfinofiniteT,mussardodifference,castrofring,esslerfiniteT,tsvelikfiniteTcorr}.
Leclair and Mussardo proposed an expansion for the one-point and two-point
functions in terms of form factors dressed by appropriate occupation
number factors containing the pseudo-energy function from the thermodynamical
Bethe Ansatz \cite{leclairmussardo}. It was shown by Saleur \cite{saleurfiniteT}
that their proposal for the two-point function is incorrect; on the
other hand, he gave a proof of the Leclair-Mussardo formula for one-point
functions provided the operator considered is the density of some
local conserved charge. His proof is based on a conjecture concerning
the expression of diagonal finite volume matrix elements in terms
of connected form factors. In view of the evidence it is now generally
accepted that the conjecture made by Leclair and Mussardo for the
one-point functions is correct; in contrast, the case of two-point
functions (and also higher ones) is not yet fully understood (see
the introductory part of section 7 for more details). Here we investigate
how finite temperature one-point functions can be expanded systematically
using finite volume $L$ as a regulator and make a proposal which
is expected to be valid for multi-point correlators as well.

Our exposition is structured as follows. In section 2, after recalling
the form factor bootstrap axioms, we present a brief review of the
approach developed in our earlier paper \cite{fftcsa} (to which we
refer the interested reader for more details), and then we state our
main results which is the description of all matrix elements containing
disconnected contributions. In Section 3 we briefly recall the two
models used for numerical comparison, which are the scaling Lee-Yang
model and the Ising model in a magnetic field. We omit the description
of the method for obtaining matrix elements from truncated conformal
space, and instead we refer the interested reader to \cite{fftcsa}
where all the necessary details can be found.

As we showed in \cite{fftcsa}, there are essentially two types of
matrix elements with disconnected contributions. Section 4 is devoted
to the first type, which is the case of diagonal matrix elements;
we present a general formula for them in terms of the symmetric evaluation
of the diagonal form factor and test it against truncated conformal
space. In section 5 we analyze diagonal matrix elements in terms of
connected form factor amplitudes, and we show that our results are
fully consistent with the above-mentioned conjecture made by Saleur
in \cite{saleurfiniteT}. In section 6 we discuss the second type
of matrix elements with disconnected contributions, namely those with
particles of exactly zero momentum in the finite volume states. Adding
the results presented in section 4 and section 6 to those obtained
in \cite{fftcsa}, we achieve a complete description of all multi-particle
matrix elements of a general local operator to all orders in $1/L$.
Section 7 is devoted to finite temperature correlation functions:
we propose a systematic method for deriving a low-temperature expansion,
which is applied to one-point functions and tested by comparing the
results to the Leclair-Mussardo expansion \cite{leclairmussardo}.
We also briefly discuss the extension of our method to the evaluation
of two-point functions. Section 8 is reserved for the conclusions.

\section{Form factors in finite volume: a brief review}

\subsection{Form factor bootstrap}

Here we give a very brief summary of the axioms of the form factor
bootstrap, because we need them in the sequel; for more details we
refer to Smirnov's review \cite{Smirnov}. Let us suppose for simplicity
that the theory has particles $A_{i}$, $i=1,\dots,N$ with masses
$m_{i}$ which are strictly non-degenerate i.e. $m_{i}\neq m_{j}$
for any $i\neq j$ (and therefore the particles are also self-conjugate).
Because of integrability, multi-particle scattering amplitudes factorize
into the product of pairwise two-particle scatterings, which are purely
elastic (in other words: diagonal). This means that any two-particle
scattering amplitude is a pure phase, which we denote by $S_{ij}\left(\theta\right)$
where $\theta$ is the relative rapidity of the incoming particles
$A_{i}$ and $A_{j}$. Incoming and outgoing asymptotic states can
be distinguished by the ordering of the rapidities:\[
|\theta_{1},\dots,\theta_{n}\rangle_{i_{1}\dots i_{n}}=\begin{cases}
|\theta_{1},\dots,\theta_{n}\rangle_{i_{1}\dots i_{n}}^{in} & :\;\theta_{1}>\theta_{2}>\dots>\theta_{n}\\
|\theta_{1},\dots,\theta_{n}\rangle_{i_{1}\dots i_{n}}^{out} & :\;\theta_{1}<\theta_{2}<\dots<\theta_{n}\end{cases}\]
and states which only differ in the order of rapidities are related
by\[
|\theta_{1},\dots,\theta_{k},\theta_{k+1},\dots,\theta_{n}\rangle_{i_{1}\dots i_{k}i_{k+1}\dots i_{n}}=S_{i_{k}i_{k+1}}(\theta_{k}-\theta_{k+1})|\theta_{1},\dots,\theta_{k+1},\theta_{k},\dots,\theta_{n}\rangle_{i_{1}\dots i_{k+1}i_{k}\dots i_{n}}\]
The normalization of these states is specified by giving the following
inner product among one-particle state:\[
\,_{j}\langle\theta^{'}|\theta\rangle_{i}=\delta_{ij}2\pi\delta(\theta^{'}-\theta)\]
For a local operator $\mathcal{O}(t,x)$ the form factors are defined
as\begin{equation}
F_{mn}^{\mathcal{O}}(\theta_{m}^{'},\dots,\theta_{1}^{'}|\theta_{1},\dots,\theta_{n})_{j_{1}\dots j_{m};i_{1}\dots i_{n}}=\,_{j_{1}\dots j_{m}}\langle\theta_{1}^{'},\dots,\theta_{m}^{'}\vert\mathcal{O}(0,0)\vert\theta_{1},\dots,\theta_{n}\rangle_{i_{1}\dots i_{n}}\label{eq:genff}\end{equation}
With the help of the crossing relations\begin{eqnarray}
 &  & F_{mn}^{\mathcal{O}}(\theta_{1}^{'},\dots,\theta_{m}^{'}|\theta_{1},\dots,\theta_{n})_{j_{1}\dots j_{m};i_{1}\dots i_{n}}=\nonumber \\
 &  & \qquad F_{m-1n+1}^{\mathcal{O}}(\theta_{1}^{'},\dots,\theta_{m-1}^{'}|\theta_{m}^{'}+i\pi,\theta_{1},\dots,\theta_{n})_{j_{1}\dots j_{m-1};j_{m}i_{1}\dots i_{n}}\nonumber \\
 &  & \qquad+\sum_{k=1}^{n}2\pi\delta_{j_{m}i_{k}}\delta(\theta_{m}^{'}-\theta_{k})\prod_{l=1}^{k-1}S_{i_{l}i_{k}}(\theta_{l}-\theta_{k})\nonumber \\
 &  & \qquad\times F_{m-1n-1}^{\mathcal{O}}(\theta_{1}^{'},\dots,\theta_{m-1}^{'}|\theta_{1},\dots,\theta_{k-1},\theta_{k+1}\dots,\theta_{n})_{j_{1}\dots j_{m-1};j_{m}i_{1}\dots i_{k-1}i_{k+1}\dots i_{n}}\label{eq:ffcrossing}\end{eqnarray}
all form factors can be expressed in terms of the elementary form
factors\begin{equation}
F_{n}^{\mathcal{O}}(\theta_{1},\dots,\theta_{n})_{i_{1}\dots i_{n}}=\langle0\vert\mathcal{O}(0,0)\vert\theta_{1},\dots,\theta_{n}\rangle_{i_{1}\dots i_{n}}\label{eq:elementaryff}\end{equation}
which satisfy the following axioms: 

I. Exchange:

\begin{center}\begin{eqnarray}
 &  & F_{n}^{\mathcal{O}}(\theta_{1},\dots,\theta_{k},\theta_{k+1},\dots,\theta_{n})_{i_{1}\dots i_{k}i_{k+1}\dots i_{n}}=\nonumber \\
 &  & \qquad S_{i_{k}i_{k+1}}(\theta_{k}-\theta_{k+1})F_{n}^{\mathcal{O}}(\theta_{1},\dots,\theta_{k+1},\theta_{k},\dots,\theta_{n})_{i_{1}\dots i_{k+1}i_{k}\dots i_{n}}\label{eq:exchangeaxiom}\end{eqnarray}
\par\end{center}

II. Cyclic permutation: \begin{equation}
F_{n}^{\mathcal{O}}(\theta_{1}+2i\pi,\theta_{2},\dots,\theta_{n})=F_{n}^{\mathcal{O}}(\theta_{2},\dots,\theta_{n},\theta_{1})\label{eq:cyclicaxiom}\end{equation}

III. Kinematical singularity\begin{equation}
-i\mathop{\textrm{Res}}_{\theta=\theta^{'}}F_{n+2}^{\mathcal{O}}(\theta+i\pi,\theta^{'},\theta_{1},\dots,\theta_{n})_{i\, j\, i_{1}\dots i_{n}}=\left(1-\delta_{i\, j}\prod_{k=1}^{n}S_{i\, i_{k}}(\theta-\theta_{k})\right)F_{n}^{\mathcal{O}}(\theta_{1},\dots,\theta_{n})_{i_{1}\dots i_{n}}\label{eq:kinematicalaxiom}\end{equation}

IV. Dynamical singularity \begin{equation}
-i\mathop{\textrm{Res}}_{\theta=\theta^{'}}F_{n+2}^{\mathcal{O}}(\theta+i\bar{u}_{jk}^{i}/2,\theta^{'}-i\bar{u}_{ik}^{j}/2,\theta_{1},\dots,\theta_{n})_{i\, j\, i_{1}\dots i_{n}}=\Gamma_{ij}^{k}F_{n+1}^{\mathcal{O}}(\theta,\theta_{1},\dots,\theta_{n})_{k\, i_{1}\dots i_{n}}\label{eq:dynamicalaxiom}\end{equation}
whenever $k$ occurs as the bound state of the particles $i$ and
$j$, corresponding to a bound state pole of the $S$ matrix of the
form\begin{equation}
S_{ij}(\theta\sim iu_{ij}^{k})\sim\frac{i\left(\Gamma_{ij}^{k}\right)^{2}}{\theta-iu_{ij}^{k}}\label{eq:Smatpole}\end{equation}
where $\Gamma_{ij}^{k}$ is the on-shell three-particle coupling and
$u_{ij}^{k}$ is the so-called fusion angle. The fusion angles satisfy\begin{eqnarray*}
m_{k}^{2} & = & m_{i}^{2}+m_{j}^{2}+2m_{i}m_{j}\cos u_{ij}^{k}\\
2\pi & = & u_{ij}^{k}+u_{ik}^{j}+u_{jk}^{i}\end{eqnarray*}
and we also used the notation $\bar{u}_{ij}^{k}=\pi-u_{ij}^{k}$.
The axioms I-IV are supplemented by the assumption of maximum analyticity
(i.e. that the form factors are meromorphic functions which only have
the singularities prescribed by the axioms) and possible further conditions
expressing properties of the particular operator whose form factors
are sought.

We remark that with the exception of free bosonic theories, all known
exact $S$ matrices satisfy \[
S_{ii}(0)=-1\]
and therefore the elementary form factors (\ref{eq:elementaryff})
have an exclusion property: they vanish whenever the rapidities of
two particles belonging to the same species coincide.

\subsection{Finite volume matrix elements to all orders in $1/L$}

Following our conventions in \cite{fftcsa}, the finite volume multi-particle
states can be denoted\[
\vert\{ I_{1},\dots,I_{n}\}\rangle_{i_{1}\dots i_{n},L}\]
where the $I_{k}$ are momentum quantum numbers and $i_{k}$ are particle
species labels. We order the momentum quantum numbers in a monotonically
decreasing sequence: $I_{n}\geq\dots\geq I_{1}$, which is just a
matter of convention. The corresponding energy levels are determined
by the Bethe-Yang equations\begin{equation}
Q_{k}(\tilde{\theta}_{1},\dots,\tilde{\theta}_{n})=m_{i_{k}}L\sinh\tilde{\theta}_{k}+\sum_{l\neq k}\delta_{i_{k}i_{l}}(\tilde{\theta}_{k}-\tilde{\theta}_{l})=2\pi I_{k}\quad,\quad k=1,\dots,n\label{eq:betheyang}\end{equation}
which must be solved with respect to the particle rapidities $\tilde{\theta}_{k}$,
where\[
\delta_{ij}(\theta)=-i\log S_{ij}(\theta)\]
are the two-particle scattering phase-shifts and the energy (with
respect to the finite volume vacuum state) can be computed as\[
\sum_{k=1}^{n}m_{i_{k}}\cosh\tilde{\theta}_{k}\]
The density of $n$-particle states can be calculated as \begin{equation}
\rho_{i_{1}\dots i_{n}}(\theta_{1},\dots,\theta_{n})=\det\mathcal{J}^{(n)}\qquad,\qquad\mathcal{J}_{kl}^{(n)}=\frac{\partial Q_{k}(\theta_{1},\dots,\theta_{n})}{\partial\theta_{l}}\quad,\quad k,l=1,\dots,n\label{eq:byjacobian}\end{equation}
We are interested in matrix elements of local operators between finite
volume multi-particle states:\[
\,_{j_{1}\dots j_{m}}\langle\{ I_{1}',\dots,I_{m}'\}\vert\mathcal{O}(0,0)\vert\{ I_{1},\dots,I_{n}\}\rangle_{i_{1}\dots i_{n},L}\]
which can be obtained numerically using truncated conformal space
(for details see \cite{fftcsa}, section 3.3). On the other hand,
using our previous results (eqn. (2.16) of \cite{fftcsa}), the finite
volume behaviour of local matrix elements can also be given as \begin{eqnarray}
 &  & \,_{j_{1}\dots j_{m}}\langle\{ I_{1}',\dots,I_{m}'\}\vert\mathcal{O}(0,0)\vert\{ I_{1},\dots,I_{n}\}\rangle_{i_{1}\dots i_{n},L}=\nonumber \\
 &  & \qquad\frac{F_{m+n}^{\mathcal{O}}(\tilde{\theta}_{m}'+i\pi,\dots,\tilde{\theta}_{1}'+i\pi,\tilde{\theta}_{1},\dots,\tilde{\theta}_{n})_{j_{m}\dots j_{1}i_{1}\dots i_{n}}}{\sqrt{\rho_{i_{1}\dots i_{n}}(\tilde{\theta}_{1},\dots,\tilde{\theta}_{n})\rho_{j_{1}\dots j_{m}}(\tilde{\theta}_{1}',\dots,\tilde{\theta}_{m}')}}+O(\mathrm{e}^{-\mu'L})\label{eq:genffrelation}\end{eqnarray}
and $\tilde{\theta}_{k}$ ($\tilde{\theta}_{k}'$) are the solutions
of the Bethe-Yang equations (\ref{eq:betheyang}) corresponding to
the state with the specified quantum numbers $I_{1},\dots,I_{n}$
($I_{1}',\dots,I_{n}'$) at the given volume $L$. The above relation
is valid provided there are no disconnected terms i.e. the left and
the right states do not contain particles with the same species and
rapidity: the sets $\left\{ (i_{1},\tilde{\theta}_{1}),\dots,(i_{n},\tilde{\theta}_{n})\right\} $
and $\left\{ (j_{1},\tilde{\theta}_{1}'),\dots,(j_{m},\tilde{\theta}_{m}')\right\} $
are disjoint. 

We recall from \cite{fftcsa} that eqns. (\ref{eq:betheyang},\ref{eq:genffrelation})
are exact to all orders of powers in $1/L$; we refer to the corrections
non-analytic in $1/L$ (eventually, as indicated, decaying exponentially)
as \emph{residual finite size effects}, following the terminology
introduced in \cite{takacspozsgay}.

\subsection{Disconnected contributions}

Let us consider a matrix element of the form\[
\,_{j_{1}\dots j_{m}}\langle\{ I_{1}',\dots,I_{m}'\}\vert\mathcal{O}(0,0)\vert\{ I_{1},\dots,I_{n}\}\rangle_{i_{1}\dots i_{n},L}\]
Disconnected terms appear when there is at least one particle in the
state on the left which occurs in the state on the right with exactly
the same rapidity. The rapidities of particles as a function of the
volume are determined by the Bethe-Yang equations (\ref{eq:betheyang})\[
Q_{k}(\tilde{\theta}_{1},\dots,\tilde{\theta}_{n})=m_{i_{k}}L\sinh\tilde{\theta}_{k}+\sum_{l\neq k}\delta_{i_{k}i_{l}}(\tilde{\theta}_{k}-\tilde{\theta}_{l})=2\pi I_{k}\quad,\quad k=1,\dots,n\]
and \[
Q_{k}(\tilde{\theta}_{1}',\dots,\tilde{\theta}_{m}')=m_{j_{k}}L\sinh\tilde{\theta}_{k}'+\sum_{l\neq k}\delta_{j_{k}j_{l}}(\tilde{\theta}_{k}'-\tilde{\theta}_{l}')=2\pi I_{k}'\quad,\quad k=1,\dots,m\]
Due to the presence of the interaction terms containing the phase
shift functions $\delta$, equality of two quantum numbers $I_{k}$
and $I_{l}'$ does not mean that the two rapidities themselves are
equal in finite volume $L$. It is easy to see that in the presence
of nontrivial scattering there are only two cases when exact equality
of the rapidities can occur:

\begin{enumerate}
\item The two states are identical, i.e. $n=m$ and \begin{eqnarray*}
\{ j_{1}\dots j_{m}\} & = & \{ i_{1}\dots i_{n}\}\\
\{ I_{1}',\dots,I_{m}'\} & = & \{ I_{1},\dots,I_{n}\}\end{eqnarray*}
In section 4 we show that the corresponding diagonal matrix element
can be written as a sum over all bipartite divisions of the set of
the $n$ particles involved (including the trivial ones when $A$
is the empty set or the complete set $\{1,\dots,n\}$)\begin{eqnarray*}
\,_{i_{1}\dots i_{n}}\langle\{ I_{1}\dots I_{n}\}|\mathcal{O}|\{ I_{1}\dots I_{n}\}\rangle_{i_{1}\dots i_{n},L} & = & \frac{1}{\rho(\{1,\dots,n\})_{L}}\times\\
 &  & \sum_{A\subset\{1,2,\dots n\}}\mathcal{F}(A)_{L}\rho(\{1,\dots,n\}\setminus A)_{L}+O(\mathrm{e}^{-\mu L})\end{eqnarray*}
where $|A|$ denotes the cardinal number (number of elements) of the
set $A$ \[
\rho(\{ k_{1},\dots,k_{r}\})_{L}=\rho_{i_{k_{1}}\dots i_{k_{r}}}(\tilde{\theta}_{k_{1}},\dots,\tilde{\theta}_{k_{r}})\]
 is the $r$-particle Bethe-Yang Jacobi determinant (\ref{eq:byjacobian})
involving only the $r$-element subset $1\leq k_{1}<\dots<k_{r}\leq n$
of the $n$ particles, and\begin{eqnarray*}
\mathcal{F}(\{ k_{1},\dots,k_{r}\})_{L} & = & F_{2r}^{s}(\tilde{\theta}_{k_{1}},\dots,\tilde{\theta}_{k_{r}})_{i_{k_{1}}\dots i_{k_{r}}}\\
F_{2l}^{s}(\theta_{1},\dots,\theta_{l})_{i_{1}\dots i_{l}} & = & \lim_{\epsilon\rightarrow0}F_{2l}^{\mathcal{O}}(\theta_{l}+i\pi+\epsilon,\dots,\theta_{1}+i\pi+\epsilon,\theta_{1},\dots,\theta_{l})_{i_{1}\dots i_{l}i_{l}\dots i_{1}}\end{eqnarray*}
is the so-called symmetric evaluation of diagonal multi-particle matrix
elements.
\item Both states are parity symmetric states in the spin zero sector, i.e.
\begin{eqnarray*}
\{ I_{1},\dots,I_{n}\} & \equiv & \{-I_{n},\dots,-I_{1}\}\\
\{ I_{1}',\dots,I'_{m}\} & \equiv & \{-I'_{m},\dots,-I'_{1}\}\end{eqnarray*}
and the particle species labels are also compatible with the symmetry,
i.e. $i_{n+1-r}=i_{r}$ and $j_{m+1-r}=j_{r}$. Furthermore, both
states must contain one (or possibly more, in a theory with more than
one species) particle of quantum number $0$, whose rapidity is then
exactly $0$ for any value of the volume $L$ due to the symmetric
assignment of quantum numbers. In section 5 we state the following
conjecture\begin{eqnarray*}
f_{2k+1,2l+1} & = & \langle\{ I_{1}',\dots,I_{k}',0,-I_{k}',\dots,-I_{1}'\}|\Phi|\{ I_{1},\dots,I_{l},0,-I_{l},\dots,-I_{1}\}\rangle_{L}\\
 & = & \frac{1}{\sqrt{\rho_{2k+1}(\tilde{\theta}_{1}',\dots,\tilde{\theta}_{k}',0,-\tilde{\theta}_{k}',\dots,-\tilde{\theta}_{1}')\rho_{2l+1}(\tilde{\theta}_{1},\dots,\tilde{\theta}_{l},0,-\tilde{\theta}_{l},\dots,-\tilde{\theta}_{1})}}\times\\
 &  & \Big(\mathcal{F}_{k,l}(\tilde{\theta}_{1}',\dots,\tilde{\theta}_{k}'|\tilde{\theta}_{1},\dots,\tilde{\theta}_{l})+mL\, F_{2k+2l}(i\pi+\tilde{\theta}_{1}',\dots,i\pi+\tilde{\theta}_{k}',\\
 &  & i\pi-\tilde{\theta}_{k}',\dots,i\pi-\tilde{\theta}_{1}',\tilde{\theta}_{1},\dots,\tilde{\theta}_{l},-\tilde{\theta}_{l},\dots,-\tilde{\theta}_{1})\Big)+O(\mathrm{e}^{-\mu L})\end{eqnarray*}
where $\rho_{n}$ is a shorthand notation for the $n$-particle Bethe-Yang
density (\ref{eq:byjacobian}) and equality is understood up to phase
conventions (cf. section 5) and \begin{eqnarray*}
 &  & \mathcal{F}_{k,l}(\theta_{1}',\dots,\theta_{k}'|\theta_{1},\dots,\theta_{l})=\\
 &  & \lim_{\epsilon\rightarrow0}F_{2k+2l+2}^{\mathcal{O}}(i\pi+\theta_{1}'+\epsilon,\dots,i\pi+\theta_{k}'+\epsilon,i\pi-\theta_{k}'+\epsilon,\dots,i\pi-\theta_{1}'+\epsilon,\\
 &  & i\pi+\epsilon,0,\theta_{1},\dots,\theta_{l},-\theta_{l},\dots,-\theta_{1})\end{eqnarray*}
is defined by assigning the same shift $\epsilon$ to all rapidities
entering the left (or equivalently the right) state and taking the
limit $\epsilon\rightarrow0$. For the sake of simplicity we assumed
above that there is a single particle species with mass $m$, but
the prescription can be easily extended to theories with more than
one particle species; an example is shown in subsection 7.2.
\end{enumerate}

\section{Exact form factors}

\subsection{Scaling Lee-Yang model}

The Hamiltonian of scaling Lee-Yang model takes the following form
in the perturbed conformal field theory framework:\[
H^{SLY}=H_{0}^{LY}+i\lambda\int_{0}^{L}dx\Phi(0,x)\]
where \[
H_{0}^{LY}=\frac{2\pi}{L}\left(L_{0}+\bar{L}_{0}-\frac{c}{12}\right)\]
is the conformal Hamiltonian and $\Phi$ is the only nontrivial primary
field, which has conformal weights $\Delta=\bar{\Delta}=-1/5$. When
$\lambda>0$ the theory above has a single particle in its spectrum
with mass $m$ that can be related to the coupling constant as \cite{lytba}
\[
\lambda=0.09704845636\dots\times m^{12/5}\]
and the bulk energy density is given by\begin{equation}
\mathcal{B}=-\frac{\sqrt{3}}{12}m^{2}\label{eq:lybulk}\end{equation}
The $S$-matrix reads \cite{CM}\begin{equation}
S_{LY}(\theta)=\frac{\sinh\theta+i\sin\frac{2\pi}{3}}{\sinh\theta-i\sin\frac{2\pi}{3}}\label{eq:Smatly}\end{equation}
and the particle occurs as a bound state of itself at $\theta=2\pi i/3$
with the three-particle coupling given by\[
\Gamma^{2}=-2\sqrt{3}\]
where the negative sign is due to the nonunitarity of the model. In
this model we define the phase-shift via the relation\[
S_{LY}(\theta)=-\mathrm{e}^{i\delta(\theta)}\]
so that $\delta(0)=0$. This means a redefinition of Bethe quantum
numbers $I_{k}$ in the Bethe-Yang equations (\ref{eq:byjacobian})
such they become half-integers for states composed of an even number
of particles; it also means that in the large volume limit, particle
momenta become\[
m\sinh\tilde{\theta}_{k}=\frac{2\pi I_{k}}{L}\]
Form factors of the trace of the stress-energy tensor $\Theta$ were
computed by Al.B. Zamolodchikov in \cite{Z1}, and using the relation\[
\Theta=i\lambda\pi(1-\Delta)\Phi\]
we can rewrite them in terms of $\Phi$. They have the form\begin{equation}
F_{n}(\theta_{1},\dots,\theta_{n})=\langle\Phi\rangle H_{n}Q_{n}(x_{1},\dots,x_{n})\prod_{i=1}^{n}\prod_{j=i+1}^{n}\frac{f(\theta_{i}-\theta_{j})}{x_{i}+x_{j}}\label{eq:lyff}\end{equation}
with the notations\begin{eqnarray*}
f(\theta) & = & \frac{\cosh\theta-1}{\cosh\theta+1/2}v(i\pi-\theta)v(i\pi+\theta)\\
v(\theta) & = & \exp\left(2\int_{0}^{\infty}dt\frac{\sinh\frac{\pi t}{2}\sinh\frac{\pi t}{3}\sinh\frac{\pi t}{6}}{t\sinh^{2}\pi t}\mathrm{e}^{i\theta t}\right)\\
x_{i} & = & \mathrm{e}^{\theta_{i}}\qquad,\qquad H_{n}=\left(\frac{3^{1/4}}{2^{1/2}v(0)}\right)^{n}\end{eqnarray*}
and the exact vacuum expectation value of the field $\Phi$ is\[
\langle\Phi\rangle=1.239394325\dots\times i\, m^{-2/5}\]
The functions $Q_{n}$ are symmetric polynomials in the variables
$x_{i}$. Defining the elementary symmetric polynomials of $n$ variables
by the relations\[
\prod_{i=1}^{n}(x+x_{i})=\sum_{i=0}^{n}x^{n-i}\sigma_{i}^{(n)}(x_{1},\dots,x_{n})\qquad,\qquad\sigma_{i}^{(n)}=0\mbox{ for }i>n\]
they can be constructed as\begin{eqnarray*}
Q_{1} & = & 1\qquad,\qquad Q_{2}=\sigma_{1}^{(2)}\qquad,\qquad Q_{3}=\sigma_{1}^{(3)}\sigma_{2}^{(3)}\\
Q_{n} & = & \sigma_{1}^{(n)}\sigma_{n-1}^{(n)}P_{n}\quad,\qquad n>3\\
P_{n} & = & \det\mathcal{M}^{(n)}\quad\mbox{where}\quad\mathcal{M}_{ij}^{(n)}=\sigma_{3i-2j+1}^{(n)}\quad,\quad i,j=1,\dots,n-3\end{eqnarray*}

\subsection{Ising model with magnetic perturbation}

The critical Ising model is the described by the conformal field theory
with $c=1/2$ and has two nontrivial primary fields: the spin operator
$\sigma$ with $\Delta_{\sigma}=\bar{\Delta}_{\sigma}=1/16$ and the
energy density $\epsilon$ with $\Delta_{\epsilon}=\bar{\Delta}_{\epsilon}=1/2$.
The magnetic perturbation, defined using the Hamiltonian (where $H_{0}^{I}$
denotes the Hamiltonian of the $c=1/2$ conformal field theory) \[
H=H_{0}^{I}+h\int_{0}^{L}dx\sigma(0,x)\]
is massive (and its physics does not depend on the sign of the external
magnetic field $h$). The spectrum and the exact $S$ matrix is described
by the famous $E_{8}$ factorized scattering theory \cite{e8}, which
contains eight particles $A_{i},\; i=1,\dots,8$ with known mass ratios,
and the mass gap relation is \cite{phonebook}\[
m_{1}=(4.40490857\dots)|h|^{8/15}\]
or\begin{equation}
h=\kappa_{h}m_{1}^{15/8}\qquad,\qquad\kappa_{h}=0.06203236\dots\label{eq:ising_massgap}\end{equation}
The bulk energy density is given by\begin{equation}
B=-0.06172858982\dots\times m^{2}\label{eq:isingbulk}\end{equation}
We also quote the scattering phase shift of two $A_{1}$ particles
for $\lambda=0$, which has the form \begin{equation}
S_{11}(\theta)=\left\{ \frac{1}{15}\right\} _{\theta}\left\{ \frac{1}{3}\right\} _{\theta}\left\{ \frac{2}{5}\right\} _{\theta}\quad,\quad\{ x\}=\frac{\sinh\theta+i\sin\pi x}{\sinh\theta-i\sin\pi x}\label{eq:s11_ising}\end{equation}
All the other amplitudes $S_{ab}$ are determined by the $S$ matrix
bootstrap \cite{e8}; we only quote the $A_{1}-A_{2}$ scattering
amplitude\[
S_{12}(\theta)=\left\{ \frac{1}{5}\right\} _{\theta}\left\{ \frac{4}{15}\right\} _{\theta}\left\{ \frac{2}{5}\right\} _{\theta}\left\{ \frac{7}{15}\right\} _{\theta}\]
because it enters some matrix elements examined later. In this model
we define the phase-shifts by the relations (for detailed explanation
cf. \cite{fftcsa})\[
S_{11}(\theta)=-\mathrm{e}^{i\delta_{11}(\theta)}\quad\mbox{and}\quad S_{12}(\theta)=\mathrm{e}^{i\delta_{12}(\theta)}\]
so that again $\delta_{11}(0)=\delta_{12}(0)=0$. The form factors
of the operator $\epsilon$ in the $E_{8}$ model were first calculated
in \cite{delfino_simonetti} and their determination was carried further
in \cite{resonances}. The exact vacuum expectation value of the field
$\epsilon$ is given by \cite{vevs}\[
\langle\epsilon\rangle=\epsilon_{h}|h|^{8/15}\qquad,\qquad\epsilon_{h}=2.00314\dots\]
or in terms of the mass scale $m=m_{1}$\[
\langle\epsilon\rangle=0.45475\dots\times m\]
For practical evaluation of form factors we used the results computed
by Delfino, Grinza and Mussardo, which can be downloaded from the
Web in \texttt{Mathematica} format \cite{isingff}. They use the following
normalized operator:\[
\Psi=\frac{\epsilon}{\langle\epsilon\rangle}\]
and so all data we plot in the sequel are understood with the same
normalization.

\section{Diagonal matrix elements}

\subsection{Form factor perturbation theory and disconnected contributions}

In the framework of conformal perturbation theory, we consider a model
with the action \begin{equation}
\mathcal{A}(\mu,\lambda)=\mathcal{A}_{\mathrm{CFT}}-\mu\int dtdx\Phi(t,x)-\lambda\int dtdx\Psi(t,x)\label{eq:nonint_Lagrangian}\end{equation}
such that in the absence of the coupling $\lambda$, the model defined
by the action $\mathcal{A}(\mu,\lambda=0)$ is integrable. The two
perturbing fields are taken as scaling fields of the ultraviolet limiting
conformal field theory, with left/right conformal weights $h_{\Phi}=\bar{h}_{\Phi}<1$
and $h_{\Psi}=\bar{h}_{\Psi}<1$, i.e. they are relevant and have
zero conformal spin, resulting in a Lorentz-invariant field theory. 

The integrable limit $\mathcal{A}(\mu,\lambda=0)$ is supposed to
define a massive spectrum, with the scale set by the dimensionful
coupling $\mu$. The exact spectrum in this case consists of some
massive particles, forming a factorized scattering theory with known
$S$ matrix amplitudes, and characterized by a mass scale $M$ (which
we take as the mass of the fundamental particle generating the bootstrap),
which is related to the coupling $\mu$ via the mass gap relation\[
\mu=\kappa M^{2-2h_{\Phi}}\]
where $\kappa$ is a (non-perturbative) dimensionless constant. 

Switching on a second independent coupling $\lambda$ in general spoils
integrability, deforms the mass spectrum and the $S$ matrix, and
in particular allows decay of the particles which are stable at the
integrable point. One way to approach the dynamics of the model is
the form factor perturbation theory proposed in \cite{nonintegrable}.
Let us denote the form factors of the operator $\Psi$ in the $\lambda=0$
theory by\[
F_{n}^{\Psi}\left(\theta_{1},\dots,\theta_{n}\right)_{i_{1}\dots i_{n}}=\langle0|\Psi(0,0)|\theta_{1}\dots\theta_{n}\rangle_{i_{1}\dots i_{n}}^{\lambda=0}\]
Using perturbation theory to first order in $\lambda$, the following
quantities can be calculated \cite{nonintegrable}:

\begin{enumerate}
\item The vacuum energy density is shifted by an amount\begin{equation}
\delta\mathcal{E}_{vac}=\lambda\left\langle 0\right|\Psi\left|0\right\rangle _{\lambda=0}.\label{vac_energy_shift}\end{equation}

\item The mass (squared) matrix $M_{ab}^{2}$ gets a correction\begin{equation}
\delta M_{ab}^{2}=2\lambda F_{2}^{\Psi}\left(i\pi\,,\,0\right)_{a\bar{b}}\delta_{m_{a},m_{b}}\label{mass_correction}\end{equation}
(where the bar denotes the antiparticle) supposing that the original
mass matrix was diagonal and of the form $M_{ab}^{2}=m_{a}^{2}\delta_{ab}\:.$
\item The scattering amplitude for the four particle process $a+b\,\rightarrow\, c+d$
is modified by \begin{equation}
\delta S_{ab}^{cd}\left(\theta,\lambda\right)=-i\lambda\frac{F_{4}^{\Psi}\left(i\pi,\,\theta+i\pi,\,0,\,\theta\right)_{\bar{c}\bar{d}ab}}{m_{a}m_{b}\sinh\theta}\quad,\quad\theta=\theta_{a}-\theta_{b}\:.\label{smatr_corr}\end{equation}
It is important to stress that the form factor amplitude in the above
expression must be defined as the so-called {}``symmetric'' evaluation\[
\lim_{\epsilon\rightarrow0}F_{4}^{\Psi}\left(i\pi+\epsilon,\,\theta+i\pi+\epsilon,\,0,\,\theta\right)_{\bar{c}\bar{d}ab}\]
(see eqn. (\ref{eq:Fs_definition}) below). It is also necessary to
keep in mind that eqn. (\ref{smatr_corr}) gives the variation of
the scattering phase when the center-of-mass energy (or, the Mandelstam
variable $s$) is kept fixed \cite{nonintegrable}. Therefore, in
terms of rapidity variables, this variation corresponds to the following:\[
\delta S_{ab}^{cd}\left(\theta,\lambda\right)=\frac{\partial S_{ab}^{cd}\left(\theta,\lambda=0\right)}{\partial\theta}\delta\theta+\lambda\left.\frac{\partial S_{ab}^{cd}\left(\theta,\lambda\right)}{\partial\lambda}\right|_{\lambda=0}\]
where \[
\delta\theta=-\frac{m_{a}\delta m_{a}+m_{a}\delta m_{a}+(m_{b}\delta m_{a}+m_{a}\delta m_{b})\cosh\theta}{m_{a}m_{b}\sinh\theta}\]
 is the shift of the rapidity variable induced by the mass corrections
given by eqn. (\ref{mass_correction}).
\end{enumerate}
It is also possible to calculate the (partial) decay width of particles
\cite{resonances}, but we do not need it here. 

We can use the above results to calculate diagonal matrix elements
involving one particle. For simplicity we present the derivation for
a theory with a single particle species. Let us start with the one-particle
case. The variation of the energy of a stationary one-particle state
with respect to the vacuum (i.e. the finite volume particle mass)
can be expressed as the difference between the first order perturbative
results for the one-particle and vacuum states in volume $L$:\begin{equation}
\Delta m(L)=\lambda L\left(\langle\{0\}|\Psi|\{0\}\rangle_{L}-\langle0|\Psi|0\rangle_{L}\right)\label{eq:finvoldm}\end{equation}
On the other hand, using Lüscher's results \cite{luscher_onept} it
only differs from the infinite volume mass in terms exponentially
falling with $L$. Using eqn. (\ref{mass_correction})\[
\Delta m(L)=\frac{\lambda}{m}F^{\Psi}(i\pi,0)+O\left(\mathrm{e}^{-\mu L}\right)\]
Similarly, the vacuum expectation value receives only corrections
falling off exponentially with $L$. Therefore we obtain\[
\langle\{0\}|\Psi|\{0\}\rangle_{L}=\frac{1}{mL}\left(F^{\Psi}(i\pi,0)+mL\langle0|\Psi|0\rangle\right)+\dots\]
with the ellipsis denoting residual finite size corrections. Note
that the factor $mL$ is just the one-particle Bethe-Yang Jacobian
$\rho_{1}(\theta)=mL\cosh\theta$ evaluated for a stationary particle
$\theta=0$. 

\begin{figure}
\begin{centering}\psfrag{fd1}{$f_{11}$}
\psfrag{l}{$l$}
\psfrag{tcsa0}{$\langle\{0\}|\Phi|\{0\}\rangle$}
\psfrag{tcsa1}{$\langle\{1\}|\Phi|\{1\}\rangle$}
\psfrag{tcsa2}{$\langle\{2\}|\Phi|\{2\}\rangle$}\includegraphics[scale=1.2]{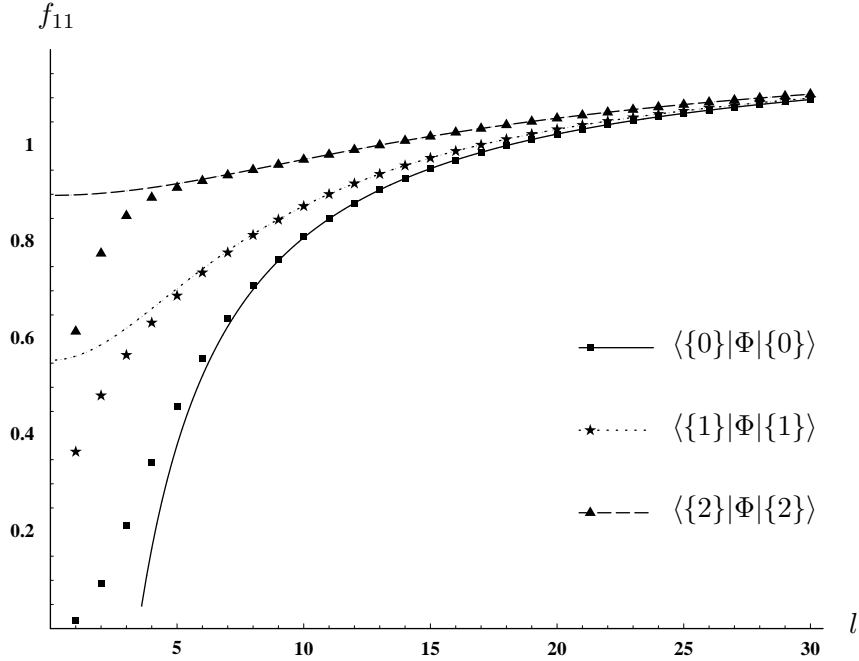}\par\end{centering}

\caption{\label{fig:d1ly}Diagonal $1$-particle matrix elements in the scaling
Lee-Yang model. The discrete points correspond to the TCSA data, while
the continuous line corresponds to the prediction from exact form
factors.}
\end{figure}

\begin{figure}
\noindent \begin{centering}\psfrag{l}{$l$}
\psfrag{ffff11}{$f_{1,1}$}
\psfrag{a0---a0}{${}_{1}\langle\{0\}|\Psi|\{0\}\rangle_{1}$}
\psfrag{a1---a1}{${}_{1}\langle\{1\}|\Psi|\{1\}\rangle_{1}$}
\psfrag{a2---a2}{${}_{1}\langle\{2\}|\Psi|\{2\}\rangle_{1}$}
\psfrag{a3---a3}{${}_{1}\langle\{3\}|\Psi|\{3\}\rangle_{1}$}\subfigure[$A_1$--$A_1$]{\includegraphics[scale=1.2]{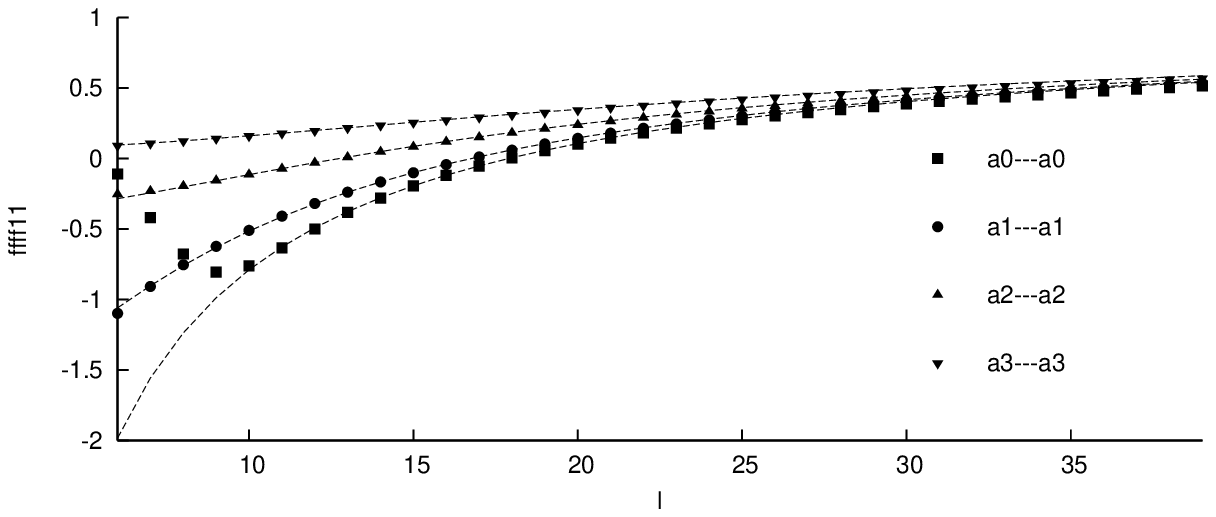}}\\
\psfrag{l}{$l$}
\psfrag{ffff22}{$f_{2,2}$}
\psfrag{b0---b0}{${}_{2}\langle\{0\}|\Psi|\{0\}\rangle_{2}$}
\psfrag{b1---b1}{${}_{2}\langle\{1\}|\Psi|\{1\}\rangle_{2}$}
\psfrag{b2---b2}{${}_{2}\langle\{2\}|\Psi|\{2\}\rangle_{2}$}
\psfrag{b3---b3}{${}_{2}\langle\{3\}|\Psi|\{3\}\rangle_{2}$}\subfigure[$A_2$--$A_2$]{\includegraphics[scale=1.2]{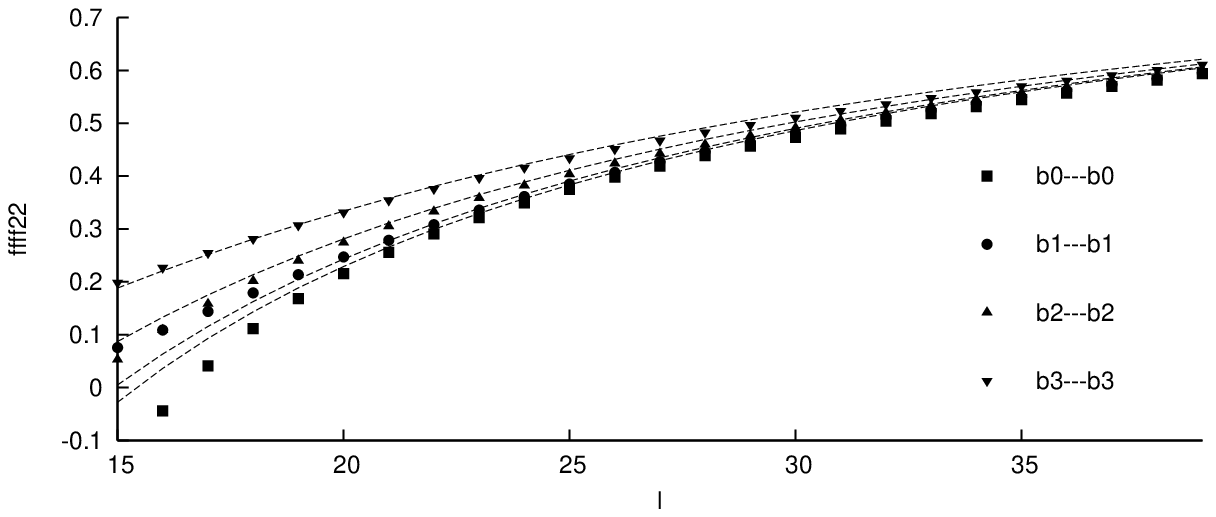}}\par\end{centering}

\caption{\label{fig:d1ising}Diagonal 1-particle matrix elements in the Ising
model. The discrete points correspond to the TCSA data, while the
continuous line corresponds to the prediction from exact form factors.}
\end{figure}

We can extend the above result to moving particles in the following
way. Up to residual finite size corrections, the one-particle energy
is given by \[
E(L)=\sqrt{m^{2}+p^{2}}\]
 with \[
p=\frac{2\pi s}{L}\]
where $s$ is the Lorentz spin (which is identical to the particle
momentum quantum number). Therefore \[
E\Delta E=m\Delta m\]
whereas perturbation theory gives: \[
\Delta E=\lambda L\left(\langle\{ s\}|\Psi|\{ s\}\rangle_{L}-\langle0|\Psi|0\rangle_{L}\right)\]
and so we obtain \begin{equation}
\langle\{ s\}|\Psi|\{ s\}\rangle_{L}=\frac{1}{\rho_{1}(\tilde{\theta})}\left(F^{\Psi}(i\pi,0)+\rho_{1}(\tilde{\theta})\langle0|\Psi|0\rangle\right)+\dots\label{eq:d1formula}\end{equation}
where \[
\sinh\tilde{\theta}=\frac{2\pi s}{mL}\,\Rightarrow\,\rho_{1}(\tilde{\theta})=\sqrt{m^{2}L^{2}+4\pi^{2}s^{2}}\]
Figure (\ref{fig:d1ly}) shows the comparison of eqn. (\ref{eq:d1formula})
to numerical data obtained from Lee-Yang TCSA: the matching is spectacular,
especially in the so-called scaling region (the volume range where
residual finite size corrections are of the order of truncation errors,
cf. \cite{fftcsa}) where the relative deviation is less than $10^{-4}$.
Here and in all following plots we use the dimensionless volume parameter
$l=mL$, and the matrix elements are also measured in units of $m$
(cf. \cite{fftcsa} for details). Diagonal one-particle matrix elements
for the Ising model are shown in figure \ref{fig:d1ising}, where
we similarly use natural units given by the mass $m=m_{1}$ of the
lightest particle $A_{1}$, just as in all subsequent plots related
to the Ising model.

\begin{figure}
\begin{centering}\psfrag{fd2}{$f_{22}$}
\psfrag{l}{$l$}
\psfrag{tcsa11}{$\langle\{\frac12,-\frac12\}|\Phi|\{\frac12,-\frac12\}\rangle$}
\psfrag{tcsa33}{$\langle\{\frac12,-\frac12\}|\Phi|\{\frac32,-\frac32\}\rangle$}
\psfrag{tcsa31}{$\langle\{\frac32,-\frac12\}|\Phi|\{\frac32,-\frac12\}\rangle$}
\psfrag{tcsa53}{$\langle\{\frac52,-\frac32\}|\Phi|\{\frac52,-\frac32\}\rangle$}\includegraphics[scale=1.2]{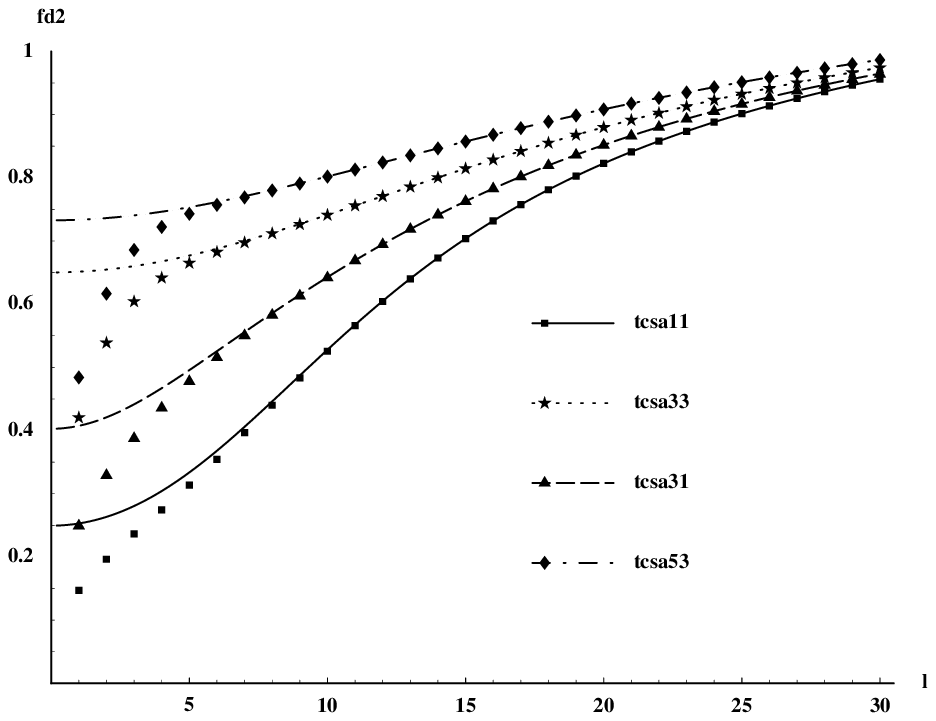}\par\end{centering}

\caption{\label{fig:d2ly}Diagonal $2$-particle matrix elements in the scaling
Lee-Yang model. The discrete points correspond to the TCSA data, while
the continuous line corresponds to the prediction from exact form
factors.}
\end{figure}

\begin{figure}
\noindent \begin{centering}\psfrag{l}{$l$}
\psfrag{ff1111}{$f_{11,11}$}
\psfrag{a-0.5a0.5---a-0.5a0.5}{${}_{11}\langle\{1/2,-1/2\}|\Psi|\{1/2,-1/2\}\rangle_{11}$}
\psfrag{a-0.5a1.5---a-0.5a1.5}{${}_{11}\langle\{3/2,-1/2\}|\Psi|\{3/2,-1/2\}\rangle_{11}$}
\psfrag{a0.5a1.5---a0.5a1.5}{${}_{11}\langle\{3/2,1/2\}|\Psi|\{3/2,1/2\}\rangle_{11}$}
\psfrag{a0.5a2.5---a0.5a2.5}{${}_{11}\langle\{5/2,1/2\}|\Psi|\{5/2,1/2\}\rangle_{11}$}\includegraphics[scale=1.1]{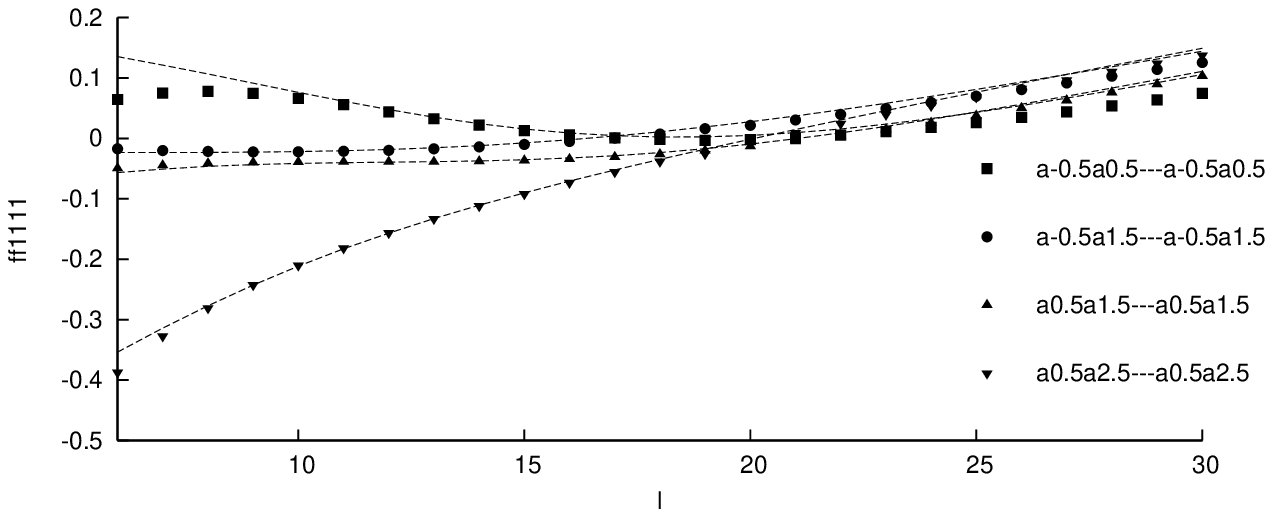}\par\end{centering}

\caption{\label{fig:d2ising}Diagonal 2-particle matrix elements in the Ising
model. The discrete points correspond to the TCSA data, while the
continuous line corresponds to the prediction from exact form factors.}
\end{figure}

One can use a similar argument to evaluate diagonal two-particle matrix
elements in finite volume. Let us assume that the theory considered
has diagonal scattering as in section 2.1. The two-particle Bethe-Yang
equations remain valid even in a non-integrable theory as long as
the total energy of the two-particle state remains under the inelastic
threshold \cite{luscher_2particle}, and therefore the energy levels
can be calculated from\begin{eqnarray*}
m_{i_{1}}L\sinh\tilde{\theta}_{1}+\delta(\tilde{\theta}_{1}-\tilde{\theta}_{2}) & = & 2\pi I_{1}\\
m_{i_{2}}L\sinh\tilde{\theta}_{2}+\delta(\tilde{\theta}_{2}-\tilde{\theta}_{1}) & = & 2\pi I_{2}\end{eqnarray*}
and (up to residual finite size corrections) \[
E_{2}(L)=E_{2pt}(L)-E_{0}(L)=m_{i_{1}}\cosh\tilde{\theta}_{1}+m_{i_{2}}\cosh\tilde{\theta}_{2}\]
where $i_{1}$ and $i_{2}$ label the particle species. After a somewhat
tedious, but elementary calculation the variation of this energy difference
with respect to $\lambda$ can be determined, using (\ref{mass_correction})
and (\ref{smatr_corr}):\begin{eqnarray*}
\Delta E_{2}(L) & = & \lambda\frac{L}{\rho_{i_{1}i_{2}}\left(\tilde{\theta}_{1},\tilde{\theta}_{2}\right)}\Big(F_{4}^{\Psi}\left(\tilde{\theta}_{2}+i\pi,\tilde{\theta}_{1}+i\pi,\tilde{\theta}_{1},\tilde{\theta}_{2}\right)_{i_{2}i_{1}i_{1}i_{2}}+m_{i_{1}}L\cosh\tilde{\theta}_{1}F_{2}^{\Psi}(i\pi,0)_{i_{2}i_{2}}\\
 &  & +m_{i_{2}}L\cosh\tilde{\theta}_{2}F^{\Psi}(i\pi,0)_{i_{1}i_{1}}\Big)\end{eqnarray*}
where all quantities (such as Bethe-Yang rapidities $\tilde{\theta}_{i}$,
masses $m_{i}$ and the two-particle state density $\rho_{2}$) are
in terms of the $\lambda=0$ theory. This result expresses the fact
that there are two sources for the variation of two-particle energy
levels: one is the mass shift of the individual particles, and the
second is due to the variation in the interaction. On the other hand,
in analogy with (\ref{eq:finvoldm}) we have\[
\Delta E_{2}(L)=\lambda L\left({}_{i_{1}i_{2}}\langle\{ I_{1},I_{2}\}|\Psi|\{ I_{1},I_{2}\}\rangle_{i_{1}i_{2},L}-\langle0|\Psi|0\rangle_{L}\right)\]
and so we obtain the following relation:\begin{eqnarray}
{}_{i_{1}i_{2}}\langle\{ I_{1},I_{2}\}|\Psi|\{ I_{1},I_{2}\}\rangle_{i_{1}i_{2},L} & = & \frac{1}{\rho_{i_{1}i_{2}}\left(\tilde{\theta}_{1},\tilde{\theta}_{2}\right)}\Big(F_{4}^{\Psi}\left(\tilde{\theta}_{2}+i\pi,\tilde{\theta}_{1}+i\pi,\tilde{\theta}_{1},\tilde{\theta}_{2}\right)_{i_{2}i_{1}i_{1}i_{2}}\nonumber \\
 &  & +m_{i_{1}}L\cosh\tilde{\theta}_{1}F_{2}^{\Psi}(i\pi,0)_{i_{2}i_{2}}\nonumber \\
 &  & +m_{i_{2}}L\cosh\tilde{\theta}_{2}F_{2}^{\Psi}(i\pi,0)_{i_{1}i_{1}}+\langle0|\Psi|0\rangle\Big)+\dots\label{eq:d2formula}\end{eqnarray}
where the ellipsis again indicate residual finite size effects. The
above argument is a generalization of the derivation of the mini-Hamiltonian
coefficient $C$ in Appendix C of \cite{takacspozsgay}. This formula
is tested against numerical data in the Lee-Yang model in figure \ref{fig:d2ly},
and the agreement is as precise as it was for the one-particle case.
Similar results can be found in the Ising case; they are shown in
figure \ref{fig:d2ising}.

\subsection{Generalization to higher number of particles}

Let us now introduce some more convenient notations. Given a state
\[
|\{ I_{1}\dots I_{n}\}\rangle_{i_{1}\dots i_{n}}\]
we denote \begin{equation}
\rho(\{ k_{1},\dots,k_{r}\})_{L}=\rho_{i_{k_{1}}\dots i_{k_{r}}}(\tilde{\theta}_{k_{1}},\dots,\tilde{\theta}_{k_{r}})\label{eq:rhonotation}\end{equation}
where $\tilde{\theta}_{l}$, $l=1,\dots,n$ are the solutions of the
$n$-particle Bethe-Yang equations (\ref{eq:betheyang}) at volume
$L$ with quantum numbers $I_{1},\dots,I_{n}$ and $\rho(\{ k_{1},\dots,k_{r}\},L)$
is the $r$-particle Bethe-Yang Jacobi determinant (\ref{eq:byjacobian})
involving only the $r$-element subset $1\leq k_{1}<\dots<k_{r}\leq n$
of the $n$ particles, evaluated with rapidities $\tilde{\theta}_{k_{1}},\dots,\tilde{\theta}_{k_{r}}$.
Let us further denote\[
\mathcal{F}(\{ k_{1},\dots,k_{r}\})_{L}=F_{2r}^{s}(\tilde{\theta}_{k_{1}},\dots,\tilde{\theta}_{k_{r}})_{i_{k_{1}}\dots i_{k_{r}}}\]
where \begin{equation}
F_{2n}^{s}(\theta_{1},\dots,\theta_{n})_{i_{1}\dots i_{n}}=\lim_{\epsilon\rightarrow0}F_{2n}^{\Psi}(\theta_{n}+i\pi+\epsilon,\dots,\theta_{1}+i\pi+\epsilon,\theta_{1},\dots,\theta_{n})_{i_{1}\dots i_{n}i_{n}\dots i_{1}}\label{eq:Fs_definition}\end{equation}
is the so-called symmetric evaluation of diagonal $n$-particle matrix
elements, which we analyze more closely in the next subsection. Note
that the exclusion property mentioned at the end of subsection 2.1
carries over to the symmetric evaluation too: (\ref{eq:Fs_definition})
vanishes whenever the rapidities of two particles of the same species
coincide. 

\begin{figure}
\begin{centering}\psfrag{fd3}{$f_{33}$}
\psfrag{l}{$l$}
\psfrag{tcsa1}{$\langle\{1,0,-1\}|\Phi|\{1,0,-1\}\rangle$}
\psfrag{tcsa2}{$\langle\{2,0,-2\}|\Phi|\{2,0,-2\}\rangle$}
\psfrag{tcsa3}{$\langle\{3,0,-3\}|\Phi|\{3,0,-3\}\rangle$}
\psfrag{tcsa4}{$\langle\{3,-1,-2\}|\Phi|\{3,-1,-2\}\rangle$}\includegraphics[scale=1.2]{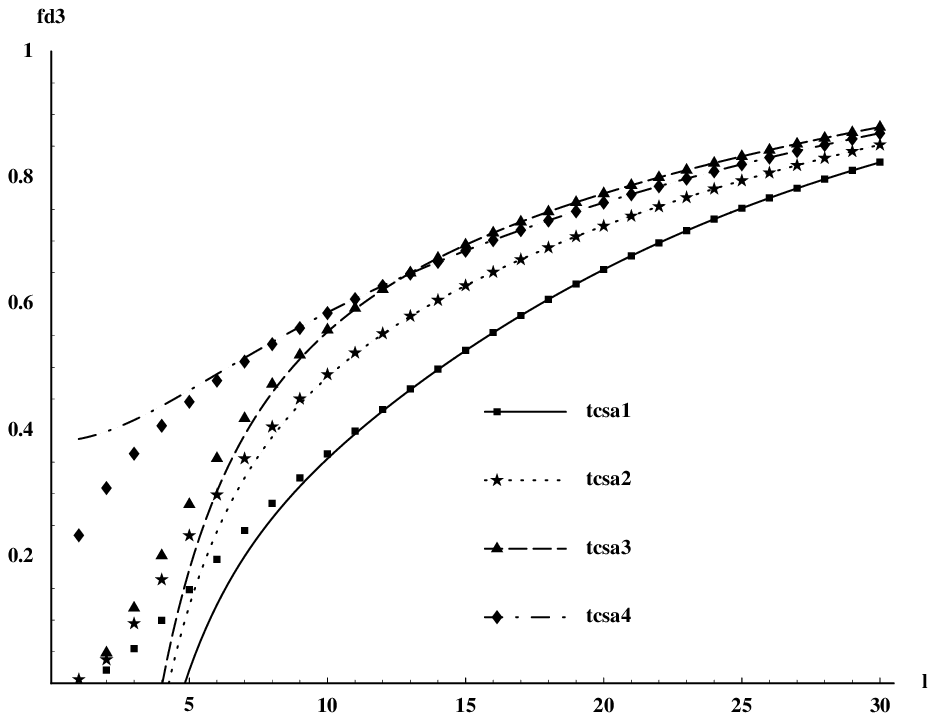}\par\end{centering}

\caption{\label{fig:d3ly}Diagonal $3$-particle matrix elements in the scaling
Lee-Yang model. The discrete points correspond to the TCSA data, while
the continuous line corresponds to the prediction from exact form
factors.}
\end{figure}

\begin{figure}
\begin{centering}\psfrag{fd4}{$f_{44}$}
\psfrag{l}{$l$}
\psfrag{tcsa1}{$\langle\{\frac32,\frac12,-\frac12,-\frac32\}|\Phi|\{\frac32,\frac12,-\frac12,-\frac32\}\rangle$}
\psfrag{tcsa2}{$\langle\{\frac52,\frac12,-\frac12,-\frac52\}|\Phi|\{\frac52,\frac12,-\frac12,-\frac52\}\rangle$}
\psfrag{tcsa3}{$\langle\{\frac72,\frac12,-\frac12,-\frac72\}|\Phi|\{\frac72,\frac12,-\frac12,-\frac72\}\rangle$}
\psfrag{tcsa4}{$\langle\{\frac72,\frac12,-\frac32,-\frac52\}|\Phi|\{\frac72,\frac12,-\frac32,-\frac52\}\rangle$}\includegraphics[scale=1.2]{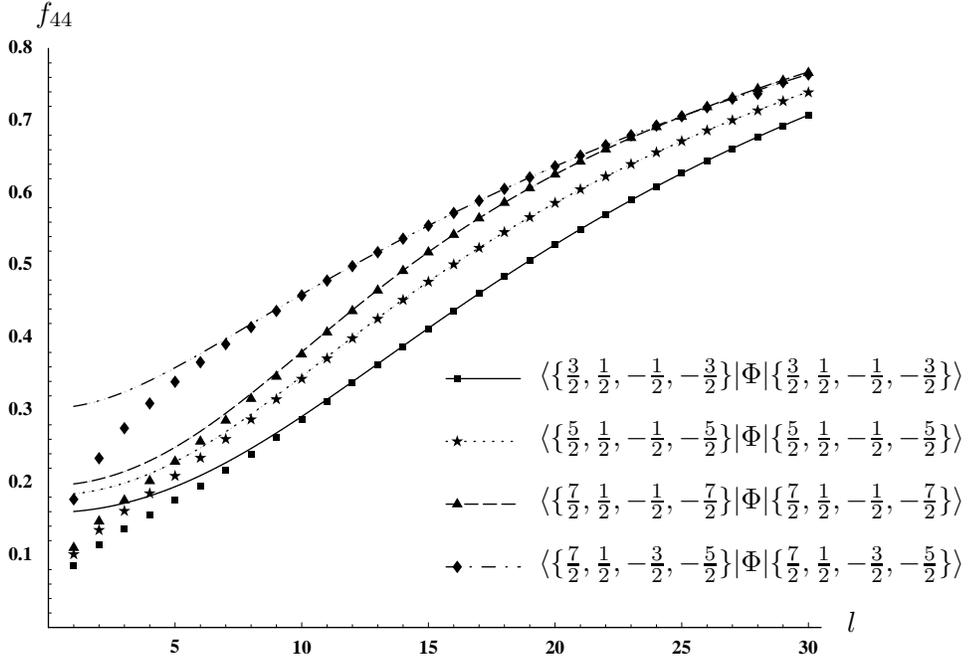}\par\end{centering}

\caption{\label{fig:d4ly}Diagonal $4$-particle matrix elements in the scaling
Lee-Yang model. The discrete points correspond to the TCSA data, while
the continuous line corresponds to the prediction from exact form
factors.}
\end{figure}

Based on the above results, we conjecture that the general rule for
a diagonal matrix element takes the form of a sum over all bipartite
divisions of the set of the $n$ particles involved (including the
trivial ones when $A$ is the empty set or the complete set $\{1,\dots,n\}$):\begin{eqnarray}
\,_{i_{1}\dots i_{n}}\langle\{ I_{1}\dots I_{n}\}|\Psi|\{ I_{1}\dots I_{n}\}\rangle_{i_{1}\dots i_{n},L} & = & \frac{1}{\rho(\{1,\dots,n\})_{L}}\times\label{eq:diaggenrule}\\
 &  & \sum_{A\subset\{1,2,\dots n\}}\mathcal{F}(A)_{L}\rho(\{1,\dots,n\}\setminus A)_{L}+O(\mathrm{e}^{-\mu L})\nonumber \end{eqnarray}
This rule can be tested against matrix elements with $n=3$ and $n=4$
in the Lee-Yang model, which are displayed in figures \ref{fig:d3ly}
and \ref{fig:d4ly}, respectively. The agreement is excellent as before,
with the relative deviation in the scaling region being of the order
of $10^{-4}$.


\section{Diagonal matrix elements in terms of connected form factors}

In this section we discuss diagonal matrix elements in terms of connected
form factors, and prove that a conjecture made by Saleur in \cite{saleurfiniteT}
exactly coincides with our eqn. (\ref{eq:diaggenrule}). To simplify
notations we omit the particle species labels; they can be restored
easily if needed.

\subsection{Relation between connected and symmetric matrix elements}

The purpose of this discussion is to give a treatment of the ambiguity
inherent in diagonal matrix elements. Due to the existence of kinematical
poles (\ref{eq:kinematicalaxiom}) the expression \[
F_{2n}(\theta_{1}+i\pi,\theta_{2}+i\pi,...,\theta_{n}+i\pi,\theta_{n},...,\theta_{2},\theta_{1})\]
which is relevant for diagonal multi-particle matrix elements, is
not well-defined. Let us consider the regularized version\[
F_{2n}(\theta_{1}+i\pi+\epsilon_{1},\theta_{2}+i\pi+\epsilon_{2},...,\theta_{n}+i\pi+\epsilon_{n},\theta_{n},...,\theta_{2},\theta_{1})\]
It was first observed in \cite{nonintegrable} that the singular parts
of this expression drop when taking the limits $\epsilon_{i}\rightarrow0$
simultaneously; however, the end result depends on the direction of
the limit, i.e. on the ratio of the $\epsilon_{i}$ parameters. The
terms that are relevant in the limit can be written in the following
general form: \begin{eqnarray}
F_{2n}(\theta_{1}+i\pi+\epsilon_{1},\theta_{2}+i\pi+\epsilon_{2},...,\theta_{n}+i\pi+\epsilon_{n},\theta_{n},...,\theta_{2},\theta_{1})=\label{mostgeneral}\\
\prod_{i=1}^{n}\frac{1}{\epsilon_{i}}\cdot\sum_{i_{1}=1}^{n}\sum_{i_{2}=1}^{n}...\sum_{i_{n}=1}^{n}a_{i_{1}i_{2}...i_{n}}(\theta_{1},\dots,\theta_{n})\epsilon_{i_{1}}\epsilon_{i_{2}}...\epsilon_{i_{n}}+\dots\nonumber \end{eqnarray}
 where $a_{i_{1}i_{2}...i_{n}}$ is a completely symmetric tensor
of rank $n$ and the ellipsis denote terms that vanish when taking
$\epsilon_{i}\rightarrow0$ simultaneously.

In our previous considerations we used the symmetric limit, which
is defined by taking all $\epsilon_{i}$ equal: \[
F_{2n}^{s}(\theta_{1},\theta_{2},...,\theta_{n})=\lim_{\epsilon\to0}F_{2n}(\theta_{1}+i\pi+\epsilon,\theta_{2}+i\pi+\epsilon,...,\theta_{n}+i\pi+\epsilon,\theta_{n},...,\theta_{2},\theta_{1})\]
It is symmetric in all the variables $\theta_{1},\dots,\theta_{n}$.
There is another evaluation with this symmetry property, namely the
so-called connected form factor, which is defined as the $\epsilon_{i}$
independent part of eqn. (\ref{mostgeneral}), i.e. the part which
does not diverge whenever any of the $\epsilon_{i}$ is taken to zero:
\begin{equation}
F_{2n}^{c}(\theta_{1},\theta_{2},...,\theta_{n})=n!\, a_{12...n}\label{eq:connected}\end{equation}
where the appearance of the factor $n!$ is simply due to the permutations
of the $\epsilon_{i}$.

\subsubsection{The relation for $n\leq3$}

We now spell out the relation between the symmetric and connected
evaluations for $n=1$, $2$ and $3$. 

The $n=1$ case is simple, since the two-particle form factor $F_{2}(\theta_{1},\theta_{2})$
has no singularities at $\theta_{1}=\theta_{2}+i\pi$ and therefore
\begin{equation}
F_{2}^{s}(\theta)=F_{2}^{c}(\theta)=F_{2}(i\pi,0)\label{eq:fs2fc2}\end{equation}
It is independent of the rapidities and will be denoted $F_{2}^{c}$
in the sequel. 

For $n=2$ we need to consider\begin{equation}
F_{4}(\theta_{1}+i\pi+\epsilon_{1},\theta_{2}+i\pi+\epsilon_{2},\theta_{2},\theta_{1})\approx\frac{a_{11}\epsilon_{1}^{2}+2a_{12}\epsilon_{1}\epsilon_{2}+a_{22}\epsilon_{2}^{2}}{\epsilon_{1}\epsilon_{2}}\label{fourresidue}\end{equation}
which gives\begin{eqnarray*}
F_{4}^{s}(\theta_{1},\theta_{2}) & = & a_{11}+2a_{12}+a_{22}\\
F_{4}^{c}(\theta_{1},\theta_{2}) & = & 2a_{12}\end{eqnarray*}
The terms $a_{11}$ and $a_{22}$ can be expressed using the two-particle
form factor. Taking an infinitesimal, but fixed $\epsilon_{2}\ne0$
\[
\mathop{\mathrm{Res}}_{\epsilon_{1}=0}F_{4}(\theta_{1}+i\pi+\epsilon_{1},\theta_{2}+i\pi+\epsilon_{2},\theta_{2},\theta_{1})=a_{22}\epsilon_{2}\]
 whereas according to (\ref{eq:dynamicalaxiom})\[
\mathop{\mathrm{Res}}_{\epsilon_{1}=0}F_{4}(\theta_{1}+i\pi+\epsilon_{1},\theta_{2}+i\pi+\epsilon_{2},\theta_{2},\theta_{1})=i\left(1-S(\theta_{1}-\theta_{2})S(\theta_{1}-\theta_{2}-i\pi-\epsilon_{2})\right)F_{2}(\theta_{2}+i\pi+\epsilon_{2},\theta_{2})\]
To first order in $\epsilon_{2}$\[
S(\theta_{1}-\theta_{2}-i\pi-\epsilon_{2})=S(\theta_{2}-\theta_{1}+\epsilon_{2})=S(\theta_{2}-\theta_{1})(1+i\varphi(\theta_{2}-\theta_{1})\epsilon_{2}+\dots)\]
where \[
\varphi(\theta)=-i\frac{d}{d\theta}\log S(\theta)\]
is the derivative of the two-particle phase shift defined before.
Therefore we obtain \[
a_{22}=\varphi(\theta_{2}-\theta_{1})F_{2}^{c}\]
and similarly \[
a_{11}=\varphi(\theta_{1}-\theta_{2})F_{2}^{c}\]
and so \begin{equation}
F_{4}^{s}(\theta_{1},\theta_{2})=F_{4}^{c}(\theta_{1},\theta_{2})+2\varphi(\theta_{1}-\theta_{2})F_{2}(i\pi,0)\label{cs_four}\end{equation}
In the case of the trace of the energy-momentum tensor $\Theta$ the
following expressions are known \cite{mussardodifference} \begin{eqnarray*}
F_{2}^{\Theta} & = & 2\pi m^{2}\\
F_{4}^{\Theta,s} & = & 8\pi m^{2}\varphi(\theta_{1}-\theta_{2})\cosh^{2}\left(\frac{\theta_{1}-\theta_{2}}{2}\right)\\
F_{4}^{\Theta,c} & = & 4\pi m^{2}\varphi(\theta_{1}-\theta_{2})\cosh(\theta_{1}-\theta_{2})\end{eqnarray*}
and they are in agreement with (\ref{cs_four}).

For $n=3$, a procedure similar to the above gives the following relation:\begin{eqnarray}
F_{6}^{s}(\theta_{1},\theta_{2},\theta_{3}) & = & F_{6}^{c}(\theta_{1},\theta_{2},\theta_{3})+\left[F_{4}^{c}(\theta_{1},\theta_{2})(\varphi(\theta_{1}-\theta_{3})+\varphi(\theta_{2}-\theta_{3}))+\mathrm{permutations}\right]\nonumber \\
 &  & +3F_{2}^{c}\left[\varphi(\theta_{1}-\theta_{2})\varphi(\theta_{1}-\theta_{3})+\mathrm{permutations}\right]\label{eq:f6sa}\end{eqnarray}
where we omitted terms that only differ by permutation of the particles.

\subsubsection{Relation between the connected and symmetric evaluation in the general
case}

Our goal is to compute the general expression \begin{equation}
F_{2n}(\theta_{1},\dots,\theta_{n}|\epsilon_{1},\dots,\epsilon_{n})=F_{2n}(\theta_{1}+i\pi+\epsilon_{1},\theta_{2}+i\pi+\epsilon_{2},...,\theta_{n}+i\pi+\epsilon_{n},\theta_{n},...,\theta_{2},\theta_{1})\label{f2n_eztkellene}\end{equation}
Let us take $n$ vertices labeled by the numbers $1,2,\dots,n$ and
let $G$ be the set of the directed graphs $G_{i}$ with the following
properties: 

\begin{itemize}
\item $G_{i}$ is tree-like. 
\item For each vertex there is at most one outgoing edge. 
\end{itemize}
For an edge going from $i$ to $j$ we use the notation $E_{ij}$.

\paragraph{Theorem 1\label{par:Theorem-1}}

(\ref{f2n_eztkellene}) can be evaluated as a sum over all graphs
in $G$, where the contribution of a graph $G_{i}$ is given by the
following two rules: 

\begin{itemize}
\item Let $A_{i}=\{ a_{1},a_{2},\dots,a_{m}\}$ be the set of vertices from
which there are no outgoing edges in $G_{i}$. The form factor associated
to $G_{i}$ is \begin{equation}
F_{2m}^{c}(\theta_{a_{1}},\theta_{a_{2}},\dots,\theta_{a_{m}})\label{egygrafformfaktora}\end{equation}
 
\item For each edge $E_{jk}$ the form factor above has to be multiplied
by \[
\frac{\epsilon_{j}}{\epsilon_{k}}\varphi(\theta_{j}-\theta_{k})\]
 
\end{itemize}
Note that since cannot contain cycles, the product of the $\epsilon_{i}/\epsilon_{j}$
factors will never be trivial (except for the empty graph with no
edges).

\paragraph*{Proof}

The proof goes by induction in $n$. For $n=1$ we have \[
F_{2}^{s}(\theta_{1})=F_{2}^{c}(\theta_{1})=F_{2}(i\pi,0)\]
This is in accordance with the theorem, because for $n=1$ there is
only the trivial graph which contains no edges and a single node. 

Now assume that the theorem is true for $n-1$ and let us take the
case of $n$ particles. Consider the residue of the matrix element
(\ref{f2n_eztkellene}) at $\epsilon_{n}=0$ while keeping all the
$\epsilon_{i}$ finite \[
R=\mathop{\mathrm{Res}}_{\epsilon_{n}=0}F_{2n}(\theta_{1}..\theta_{n}|\epsilon_{1}..\epsilon_{n})\]
According to the theorem the graphs contributing to this residue are
exactly those for which the vertex $n$ has an outgoing edge and no
incoming edges. Let $R_{j}$ be sum of the diagrams where the outgoing
edge is $E_{nj}$ for some $j=1,\dots,n-1$, and so \[
R=\sum_{j=1}^{n-1}R_{j}\]
The form factors appearing in $R_{j}$ do not depend on $\theta_{n}$.
Therefore we get exactly the diagrams that are needed to evaluate
$F_{2(n-1)}(\theta_{1}..\theta_{n-1}|\epsilon_{1}..\epsilon_{n-1})$,
apart from the proportionality factor associated to the link $E_{nj}$
and so \[
R_{j}=\frac{\epsilon_{j}}{\epsilon_{n}}\varphi(\theta_{j}-\theta_{n})F_{2(n-1)}(\theta_{1}..\theta_{n-1}|\epsilon_{1}..\epsilon_{n-1})\]
and summing over $j$ gives \begin{equation}
R=(\epsilon_{1}\varphi(\theta_{1}-\theta_{n})+\epsilon_{2}\varphi(\theta_{2}-\theta_{n})+\dots+\epsilon_{n-1}\varphi(\theta_{n-1}-\theta_{n}))F_{2(n-1)}(\theta_{1}..\theta_{n-1}|\epsilon_{1}..\epsilon_{n-1})\label{erremitlepsz}\end{equation}
In order to prove the theorem, we only need to show that the residue
indeed takes this form. On the other hand, the kinematical residue
axiom (\ref{eq:kinematicalaxiom}) gives \[
R=i\left(1-\prod_{j=1}^{n-1}S(\theta_{n}-\theta_{j})S(\theta_{n}-\theta_{j}-i\pi-\epsilon_{j})\right)F_{2(n-1)}(\theta_{1}..\theta_{n-1}|\epsilon_{1}..\epsilon_{n-1})\]
 which is exactly the same as eqn. (\ref{erremitlepsz}) when expanded
to first order in $\epsilon_{j}$.

We thus checked that the theorem gives the correct result for the
terms that include a $1/\epsilon_{n}$ singularity. Using symmetry
in the rapidity variables this is true for all the terms that include
at least one $1/\epsilon_{i}$ for an arbitrary $i$. There is only
one diagram that cannot be generated by the inductive procedure, namely
the empty graph. However, there are no singularities ($1/\epsilon_{i}$
factors) associated to it, and it gives $F_{2n}^{c}(\theta_{1},\dots,\theta_{n})$
by definition. \emph{Qed}.

\begin{figure}
\noindent \begin{centering}\includegraphics[scale=1.2]{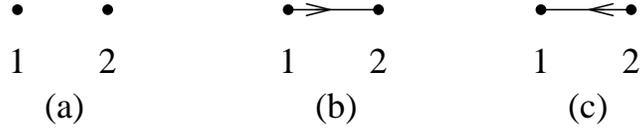}\par\end{centering}

\caption{\label{fig:neq2gr} The graphs relevant for $n=2$}
\end{figure}

\begin{figure}
\noindent \begin{centering}\includegraphics[scale=1.2]{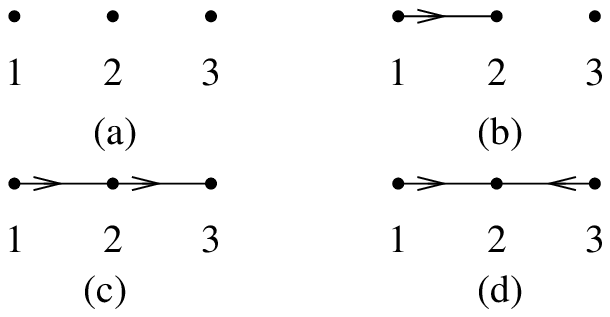}\par\end{centering}

\caption{\label{fig:neq3gr} The graphs relevant for $n=3$}
\end{figure}

We now illustrate how the theorem works. For $n=2$, there are only
three graphs, depicted in figure \ref{fig:neq2gr}. Applying the rules
yields \[
F_{4}(\theta_{1},\theta_{2}|\epsilon_{1},\epsilon_{2})=F_{4}^{c}(\theta_{1},\theta_{2})+\varphi(\theta_{1}-\theta_{2})\left(\frac{\epsilon_{1}}{\epsilon_{2}}+\frac{\epsilon_{2}}{\epsilon_{1}}\right)F_{2}^{c}\]
which gives back (\ref{cs_four}) upon putting $\epsilon_{1}=\epsilon_{2}$.
For $n=3$ there are $4$ different kinds of graphs, the representatives
of which are shown in figure \ref{fig:neq3gr}; all other graphs can
be obtained by permuting the node labels $1,2,3$. The contributions
of these graphs are

\begin{eqnarray*}
(a) & : & F_{6}^{c}(\theta_{1},\theta_{2},\theta_{3})\\
(b) & : & \frac{\epsilon_{2}}{\epsilon_{1}}\varphi(\theta_{1}-\theta_{2})F_{4}^{c}(\theta_{2},\theta_{3})\\
(c) & : & \frac{\epsilon_{2}}{\epsilon_{1}}\frac{\epsilon_{3}}{\epsilon_{2}}\varphi(\theta_{1}-\theta_{2})\varphi(\theta_{2}-\theta_{3})F_{2}^{c}=\frac{\epsilon_{3}}{\epsilon_{1}}\varphi(\theta_{1}-\theta_{2})\varphi(\theta_{2}-\theta_{3})F_{2}^{c}\\
(d) & : & \frac{\epsilon_{2}}{\epsilon_{1}}\frac{\epsilon_{2}}{\epsilon_{3}}\varphi(\theta_{1}-\theta_{2})\varphi(\theta_{3}-\theta_{2})F_{2}^{c}\end{eqnarray*}
Adding up all the contributions and putting $\epsilon_{1}=\epsilon_{2}=\epsilon_{3}$
we recover eqn. (\ref{eq:f6sa}).

\subsection{Consistency with Saleur's proposal}

Saleur proposed an expression for diagonal matrix elements in terms
of connected form factors in \cite{saleurfiniteT}, which is partially
based on earlier work by Balog \cite{balogtba} and also on the determinant
formula for normalization of states in the framework of algebraic
Bethe Ansatz, derived by Gaudin, and also by Korepin (see \cite{qism}
and references therein). To describe it, we must extend the normalization
of finite volume states defined in \cite{fftcsa} to the case when
the particle rapidities form a proper subset of some multi-particle
Bethe-Yang solution. 

According to \cite{fftcsa}, the normalization of a finite volume
state is given by\[
\vert\{ I_{1},\dots,I_{n}\}\rangle_{L}=\frac{1}{\sqrt{\rho_{n}(\tilde{\theta}_{1},\dots,\tilde{\theta}_{n})}}\vert\tilde{\theta}_{1},\dots,\tilde{\theta}_{n}\rangle\]
in terms of the infinite volume state with rapidities $\tilde{\theta}_{1},\dots,\tilde{\theta}_{n}$,
which are the solutions of the Bethe-Yang equations (\ref{eq:betheyang})
for the given quantum numbers $I_{1},\dots,I_{n}$ at volume $L$
(we again omit the particle species labels, and also denote the $n$-particle
determinant by $\rho_{n}$). Let us take a subset of particle indices
$A\in\{1,\dots,n\}$ and define the corresponding sub-determinant
by\[
\tilde{\rho}_{n}(\tilde{\theta}_{1},\dots,\tilde{\theta}_{n}|A)=\det\mathcal{J}_{A}^{(n)}\]
where $\mathcal{J}_{A}^{(n)}$ is the sub-matrix of the matrix $\mathcal{J}^{(n)}$
defined in eqn. (\ref{eq:byjacobian}) which is given by choosing
the elements whose indices belong to $A$. The full matrix can be
written explicitly as\[
\mathcal{J}^{(n)}=\begin{pmatrix}E_{1}L+\varphi_{12}+\dots+\varphi_{1n} & -\varphi_{12} & \dots & -\varphi_{1n}\\
-\varphi_{12} & E_{2}L+\varphi_{21}+\varphi_{23}+\dots+\varphi_{2n} & \dots & -\varphi_{2n}\\
\vdots & \vdots & \ddots & \vdots\\
-\varphi_{1n} & -\varphi_{2n} & \dots & E_{n}L+\varphi_{1n}+\dots+\varphi_{n-1,n}\end{pmatrix}\]
where the following abbreviations were used: $E_{i}=m_{i}\cosh\theta_{i}$,
$\varphi_{ij}=\varphi_{ji}=\varphi(\theta_{i}-\theta_{j})$. Note
that $\tilde{\rho}_{n}$ depends on all the rapidities, not just those
which correspond to elements of $A$. It is obvious that\[
\rho_{n}(\tilde{\theta}_{1},\dots,\tilde{\theta}_{n})\equiv\tilde{\rho}_{n}(\tilde{\theta}_{1},\dots,\tilde{\theta}_{n}|\{1,\dots,n\})\]
Saleur proposed the definition\begin{equation}
\langle\{\tilde{\theta}_{k}\}_{k\in A}\vert\{\tilde{\theta}_{k}\}_{k\in A}\rangle_{L}=\tilde{\rho}_{n}(\tilde{\theta}_{1},\dots,\tilde{\theta}_{n}|A)\label{eq:detnorm}\end{equation}
where \[
\vert\{\tilde{\theta}_{k}\}_{k\in A}\rangle_{L}\]
is a {}``partial state'' which contains only the particles with
index in $A$, but with rapidities that solve the Bethe-Yang equations
for the full $n$-particle state. Note that this is not a proper state
in the sense that it is not an eigenstate of the Hamiltonian since
the particle rapidities do not solve the Bethe-Yang equations relevant
for a state consisting of $|A|$ particles (where $|A|$ denotes the
cardinal number -- i.e. number of elements -- of the set $A$). The
idea behind this proposal is that the density of these partial states
in rapidity space depends on the presence of the other particles which
are not included, and indeed it is easy to see that it is given by
$\tilde{\rho}_{n}(\tilde{\theta}_{1},\dots,\tilde{\theta}_{n}|A)$.

In terms of the above definitions, Saleur's conjecture for the diagonal
matrix element is\begin{eqnarray}
\,_{i_{1}\dots i_{n}}\langle\{ I_{1}\dots I_{n}\}|\Psi|\{ I_{1}\dots I_{n}\}\rangle_{i_{1}\dots i_{n},L} & = & \frac{1}{\rho_{n}(\tilde{\theta}_{1},\dots,\tilde{\theta}_{n})}\times\label{eq:diaggenrulesaleur}\\
 &  & \sum_{A\subset\{1,2,\dots n\}}F_{2|A|}^{c}(\{\tilde{\theta}_{k}\}_{k\in A})\tilde{\rho}(\tilde{\theta}_{1},\dots,\tilde{\theta}_{n}|A)+O(\mathrm{e}^{-\mu L})\nonumber \end{eqnarray}
which is just the standard representation of the full matrix element
as the sum of all the connected contributions provided we accept eqn.
(\ref{eq:detnorm}). The full amplitude is obtained by summing over
all possible bipartite divisions of the particles, where the division
is into particles that are connected to the local operator, giving
the connected form factor $F^{c}$ and into those that simply go directly
from the initial to the final state which contribute the norm of the
corresponding partial multi-particle state. 

Using the results of subsection 5.1, it is easy to check explicitly
(which we did up to $n=3$) that our rule for the diagonal matrix
elements as given in eqn. (\ref{eq:diaggenrule}) is equivalent to
eqn. (\ref{eq:diaggenrulesaleur}). We now give a complete proof for
the general case.

\paragraph{Theorem 2}

\begin{equation}
\sum_{A\subset N}F_{2|A|}^{c}(\{\theta_{k}\}_{k\in A})\tilde{\rho}(\theta_{1},\dots,\theta_{n}|A)=\sum_{A\subset N}F_{2|A|}^{s}(\{\theta_{k}\}_{k\in A})\rho(\{\theta_{k}\}_{k\in N\setminus A})\label{ketfelekepp}\end{equation}
where we denoted $N=\{1,2,\dots,n\}$.

\paragraph{Proof}

The two sides of eqn. (\ref{ketfelekepp}) differ in two ways: 

\begin{itemize}
\item The form factors on the right hand side are evaluated according to
the ,,symmetric'' prescription, and in addition to the connected
part also they contain extra terms, which are proportional to connected
form factors with fewer particles. 
\item The densities $\tilde{\rho}$ on the left hand side are not determinants
of the form (\ref{eq:byjacobian}) written down in terms of the particles
contained in $N\setminus A$: they contain additional terms due to
the presence of the particles in $A$ as well. 
\end{itemize}
Here we show that eqn. (\ref{ketfelekepp}) is merely a reorganization
of these terms. 

For simplicity consider first the term on the left hand side which
corresponds to $A=\{ m+1,m+2,\dots,n\}$, i.e. \[
F_{2m}^{c}(\theta_{m+1},\dots,\theta_{n})\tilde{\rho}(\theta_{1},\dots,\theta_{n}|A)\]
We expand $\tilde{\rho}$ in terms of the physical multi-particle
densities $\rho$. In order to accomplish this, it is useful to rewrite
the sub-matrix $\mathcal{J}_{N\setminus A}^{n}{}$ as \[
\mathcal{J}^{(n)}|_{N\setminus A}=\mathcal{J}^{m}(\theta_{1},\dots,\theta_{m})+\begin{pmatrix}\sum\limits _{i=m+1}^{n}\varphi_{1i}\\
 &  & \sum\limits _{i=m+1}^{n}\varphi_{2i}\\
 &  &  &  & \ddots\\
 &  &  &  &  &  & \sum\limits _{i=m+1}^{n}\varphi_{mi}\end{pmatrix}\]
where $\mathcal{J}^{m}$ is the $m$-particle Jacobian matrix which
does not contain any terms depending on the particles in $A$. The
determinant of $\mathcal{J}_{N\setminus A}^{n}{}$ can be written
as a sum over the subsets of $N\setminus A$. For a general subset
$B\subset N\setminus A$ let us use the notation $B=\{ b_{1},b_{2},\dots,b_{|B|}\}$.
We can then write \begin{equation}
\tilde{\rho}(\theta_{1},\dots,\theta_{n}|A)=\text{det}\mathcal{J}^{(n)}|_{N\setminus A}=\sum_{B}\left[\rho(N\setminus(A\cup B))\prod_{i=1}^{|B|}\left(\sum_{c_{i}=m+1}^{n}\varphi_{b_{i},c_{i}}\right)\right]\label{egyfajta}\end{equation}
 where $\rho(N\setminus(A\cup B))$ is the $\rho$-density (\ref{eq:byjacobian})
written down with the particles in $N\setminus(A\cup B)$. 

Applying a suitable permutation of variables we can generalize eqn.
(\ref{egyfajta}) to an arbitrary subset $A\subset N$: \begin{equation}
\tilde{\rho}(\theta_{1},\dots,\theta_{n}|A)=\text{det}\mathcal{J}^{(n)}|_{N\setminus A}=\sum_{B}\rho(N\setminus(A\cup B))\sum_{C}(\prod_{i=1}^{|B|}\varphi_{b_{i},c_{i}})\label{egyfajta2}\end{equation}
where the second summation goes over all the sets $C=\{ c_{1},c_{2},\dots,c_{|B|}\}$
with $|C|=|B|$ and $c_{i}\in A$. The left hand side of eqn. (\ref{ketfelekepp})
can thus be written as \begin{eqnarray}
\sum_{A\subset N}F_{2|A|}^{c}(\{\theta_{k}\}_{k\in A})\tilde{\rho}(\theta_{1},\dots,\theta_{n}|A) & = & \sum_{\begin{array}{c}
A,B\subset N\\
A\cap B=\emptyset\end{array}}\rho(N\setminus(A\cup B))\sum_{C}F_{(A,B,C)}\label{ketfelekepp2}\\
\mbox{where} &  & F_{(A,B,C)}=F_{2|A|}^{c}(\{\theta_{k}\}_{k\in A})\prod_{i=1}^{|B|}\varphi_{b_{i},c_{i}}\nonumber \end{eqnarray}
We now show that there is a one-to-one correspondence between all
the terms in (\ref{ketfelekepp2}) and those on the right hand side
of (\ref{ketfelekepp}) if the symmetric evaluations $F_{2k}^{s}$
are expanded according to Theorem 1. To each triplet $(A,B,C)$ let
us assign the graph $G_{(A,B,C)}$ defined as follows: 

\begin{itemize}
\item The vertices of the graph are the elements of the set $A\cup B$. 
\item There are exactly $|B|$ edges in the graph, which start at $b_{i}$
and end at $c_{i}$ with $i=1,\dots,|B|$. 
\end{itemize}
The contribution of $G_{(A,B,C)}$ to $F_{2(|A|+|B|)}^{s}(\{\theta_{k}\}_{k\in A\cup B})$
is nothing else than $F_{(A,B,C)}$ which can be proved by applying
the rules of Theorem 1. Note that all the possible diagrams with at
most $n$ vertices are contained in the above list of the $G_{(A,B,C)}$,
because a general graph $G$ satisfying \textit{\emph{the conditions}}
in Theorem 1 can be characterized by writing down the set of vertices
with and without outgoing edges (in this case $B$ and $A$) and the
endpoints of the edges (in this case $C$).

It is easy to see that the factors $\rho(N\setminus(A\cup B))$ multiplying
the $F_{(A,B,C)}$ in (\ref{ketfelekepp2}) are also the correct ones:
they are just the density factors multiplying $F_{2(|A|+|B|)}^{s}(\{\theta_{k}\}_{k\in A\cup B})$
on the right hand side of (\ref{ketfelekepp}). \emph{Qed}.

\section{Zero-momentum particles}

\subsection{Scaling Lee-Yang model}

In the scaling Lee-Yang model, with a single type of particle, there
can only be a single particle of zero momentum in a multi-particle
state due to the exclusion principle. For the momentum to be exactly
zero in finite volume it is necessary that the all other particles
should come with quantum numbers in pairs of opposite sign, which
means that the state must have $2n+1$ particles in a configuration\[
|\{ I_{1},\dots,I_{n},0,-I_{n},\dots,-I_{1}\}\rangle_{L}\]
Therefore we consider matrix elements of the form\[
\langle\{ I_{1}',\dots,I_{k}',0,-I_{k}',\dots,-I_{1}'\}|\Phi|\{ I_{1},\dots,I_{l},0,-I_{l},\dots,-I_{1}\}\rangle_{L}\]
(with $k=0$ or $l=0$ corresponding to a state containing a single
stationary particle). We also suppose that the two sets $\{ I_{1},\dots,I_{k}\}$
and $\{ I_{1}',\dots,I_{l}'\}$ are not identical, otherwise we have
the case of diagonal matrix elements treated in section 4.

We need to examine form factors of the form\[
F_{2k+2l+2}(i\pi+\theta_{1}',\dots,i\pi+\theta_{k}',i\pi-\theta_{k}',\dots,i\pi-\theta_{1}',i\pi+\theta,0,\theta_{1},\dots,\theta_{l},-\theta_{l},\dots,-\theta_{1})\]
where the particular ordering of the rapidities was chosen to ensure
that no additional $S$ matrix factors appear in the disconnected
terms of the crossing relation (\ref{eq:ffcrossing}). Using the singularity
axiom (\ref{eq:kinematicalaxiom}), plus unitarity and crossing symmetry
of the $S$-matrix it is easy to see that the residue of the above
function at $\theta=0$ vanishes, and so it has a finite limit as
$\theta\rightarrow0$. However, this limit depends on direction just
as in the case of the diagonal matrix elements considered in section
4. Therefore we must specify the way it is taken, and just as previously
we use a prescription that is maximally symmetric in all variables:
we choose to shift all rapidities entering the left hand state with
the same amount to define\begin{eqnarray}
 &  & \mathcal{F}_{k,l}(\theta_{1}',\dots,\theta_{k}'|\theta_{1},\dots,\theta_{l})=\nonumber \\
 &  & \lim_{\epsilon\rightarrow0}F_{2k+2l+2}(i\pi+\theta_{1}'+\epsilon,\dots,i\pi+\theta_{k}'+\epsilon,i\pi-\theta_{k}'+\epsilon,\dots,i\pi-\theta_{1}'+\epsilon,\nonumber \\
 &  & i\pi+\epsilon,0,\theta_{1},\dots,\theta_{l},-\theta_{l},\dots,-\theta_{1})\label{eq:oddoddlimitdef}\end{eqnarray}
Using the above definition, by analogy to (\ref{eq:diaggenrule})
we conjecture that\begin{eqnarray}
f_{2k+1,2l+1} & = & \langle\{ I_{1}',\dots,I_{k}',0,-I_{k}',\dots,-I_{1}'\}|\Phi|\{ I_{1},\dots,I_{l},0,-I_{l},\dots,-I_{1}\}\rangle_{L}\label{eq:oddoddlyrule}\\
 & = & \frac{1}{\sqrt{\rho_{2k+1}(\tilde{\theta}_{1}',\dots,\tilde{\theta}_{k}',0,-\tilde{\theta}_{k}',\dots,-\tilde{\theta}_{1}')\rho_{2l+1}(\tilde{\theta}_{1},\dots,\tilde{\theta}_{l},0,-\tilde{\theta}_{l},\dots,-\tilde{\theta}_{1})}}\times\nonumber \\
 &  & \Big(\mathcal{F}_{k,l}(\tilde{\theta}_{1}',\dots,\tilde{\theta}_{k}'|\tilde{\theta}_{1},\dots,\tilde{\theta}_{l})+mL\, F_{2k+2l}(i\pi+\tilde{\theta}_{1}',\dots,i\pi+\tilde{\theta}_{k}',\nonumber \\
 &  & i\pi-\tilde{\theta}_{k}',\dots,i\pi-\tilde{\theta}_{1}',\tilde{\theta}_{1},\dots,\tilde{\theta}_{l},-\tilde{\theta}_{l},\dots,-\tilde{\theta}_{1})\Big)+O(\mathrm{e}^{-\mu L})\nonumber \end{eqnarray}
where $\tilde{\theta}$ denote the solutions of the appropriate Bethe-Yang
equations at volume $L$, $\rho_{n}$ is a shorthand notation for
the $n$-particle Bethe-Yang density (\ref{eq:byjacobian}) and equality
is understood up to phase factors. We recall from our previous work
\cite{fftcsa} that relative phases of multi-particle states are in
general fixed differently in the form factor bootstrap and TCSA. Also
note that reordering particles gives phase factors on the right hand
side according to the exchange axiom (\ref{eq:exchangeaxiom}). This
issue is obviously absent in the case of diagonal matrix elements
treated in sections 4 and 5, since any such phase factor cancels out
between the state and its conjugate. Such phases do not affect correlation
functions, or as a consequence, any physically relevant quantities
since they can all be expressed in terms of correlators. 

There is some argument that can be given in support of eqn. (\ref{eq:oddoddlyrule}).
Note that the zero-momentum particle occurs in both the left and right
states, which actually makes it unclear how to define a density similar
to $\tilde{\rho}$ in (\ref{eq:detnorm}). Such a density would take
into account the interaction with the other particles. However, the
nonzero rapidities entering of the two states are different and therefore
there is no straightforward way to apply Saleur's recipe (\ref{eq:diaggenrulesaleur})
here. Using the maximally symmetric definition (\ref{eq:oddoddlimitdef})
the shift $\epsilon$ can be equally put on the right hand side rapidities
as well, and therefore we expect that the density factor multiplying
the term $F_{2k+2l}$ in (\ref{eq:oddoddlyrule}) would be the one-particle
state density in which none of the other rapidities appear, which
is exactly $mL$ for a stationary particle. This is a natural guess
from eqn. (\ref{eq:diaggenrule}) which states that when diagonal
matrix elements are expressed using the symmetric evaluation, only
densities of the type $\rho$ appear. 

Another argument can be formulated using the observation that eqn.
(\ref{eq:oddoddlyrule}) is only valid if $\mathcal{F}_{k,l}$ is
defined as in (\ref{eq:oddoddlimitdef}); all other possible ways
to take the limit can be related in a simple way to this definition
and so the rule (\ref{eq:oddoddlyrule}) can be rewritten appropriately.
Let us consider two other natural choices\begin{eqnarray*}
 &  & \mathcal{F}_{k,l}^{+}(\theta_{1}',\dots,\theta_{k}'|\theta_{1},\dots,\theta_{l})=\\
 &  & \lim_{\epsilon\rightarrow0}F_{2k+2l+2}(i\pi+\theta_{1}',\dots,i\pi+\theta_{k}',i\pi-\theta_{k}',\dots,i\pi-\theta_{1}',i\pi,\epsilon,\theta_{1},\dots,\theta_{l},-\theta_{l},\dots,-\theta_{1})\\
 &  & \mathcal{F}_{k,l}^{-}(\theta_{1}',\dots,\theta_{k}'|\theta_{1},\dots,\theta_{l})=\\
 &  & \lim_{\epsilon\rightarrow0}F_{2k+2l+2}(i\pi+\theta_{1}',\dots,i\pi+\theta_{k}',i\pi-\theta_{k}',\dots,i\pi-\theta_{1}',i\pi+\epsilon,0,\theta_{1},\dots,\theta_{l},-\theta_{l},\dots,-\theta_{1})\end{eqnarray*}
in which the shift is put only on the zero-momentum particle on the
right/left, respectively. Using the kinematical residue axiom (\ref{eq:kinematicalaxiom}),
$\mathcal{F}^{\pm}$ can be related to $\mathcal{F}$ via\begin{eqnarray*}
 &  & \mathcal{F}_{k,l}(\theta_{1}',\dots,\theta_{k}'|\theta_{1},\dots,\theta_{l})=\mathcal{F}_{k,l}^{+}(\theta_{1}',\dots,\theta_{k}'|\theta_{1},\dots,\theta_{l})\\
 &  & +2\sum_{i=1}^{l}\varphi(\theta_{i})F_{2k+2l}(i\pi+\theta_{1}',\dots,i\pi+\theta_{k}',i\pi-\theta_{k}',\dots,i\pi-\theta_{1}',\theta_{1},\dots,\theta_{l},-\theta_{l},\dots,-\theta_{1})\\
 &  & \mathcal{F}_{k,l}(\theta_{1}',\dots,\theta_{k}'|\theta_{1},\dots,\theta_{l})=\mathcal{F}_{k,l}^{-}(\theta_{1}',\dots,\theta_{k}'|\theta_{1},\dots,\theta_{l})\\
 &  & -2\sum_{i=1}^{k}\varphi(\theta_{i}')F_{2k+2l}(i\pi+\theta_{1}',\dots,i\pi+\theta_{k}',i\pi-\theta_{k}',\dots,i\pi-\theta_{1}',\theta_{1},\dots,\theta_{l},-\theta_{l},\dots,-\theta_{1})\end{eqnarray*}
With the help of the above relations eqn. (\ref{eq:oddoddlyrule})
can also be rewritten in terms of $\mathcal{F}^{\pm}$. The way $\mathcal{F}$
and therefore also eqn. (\ref{eq:oddoddlyrule}) are expressed in
terms of $\mathcal{F^{\pm}}$ shows a remarkable and natural symmetry
under the exchange of the left and right state (and correspondingly
$\mathcal{F}^{+}$ with $\mathcal{F}^{-}$), which provides a further
support to our conjecture. 

\begin{figure}
\begin{centering}\psfrag{fd13}{$|f_{13}|$}
\psfrag{l}{$l$}
\psfrag{tcsa1}{$\langle\{0\}|\Phi|\{1,0,-1\}\rangle$}
\psfrag{tcsa2}{$\langle\{0\}|\Phi|\{2,0,-2\}\rangle$}
\psfrag{tcsa3}{$\langle\{0\}|\Phi|\{3,0,-3\}\rangle$}\includegraphics[scale=1.2]{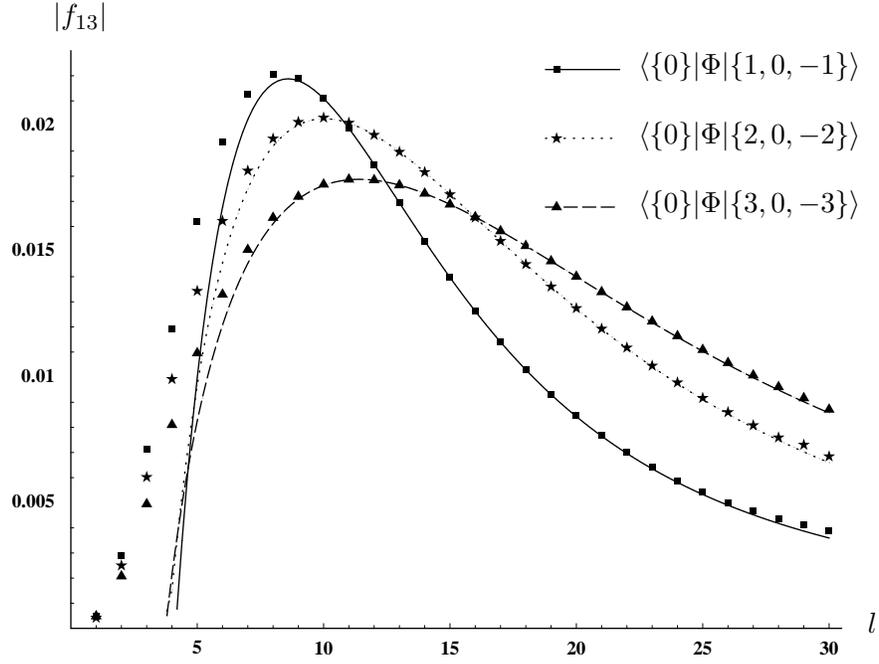}\par\end{centering}

\caption{\label{fig:fd13ly}$1$-particle--$3$-particle matrix elements in
the scaling Lee-Yang model. The discrete points correspond to the
TCSA data, while the continuous line corresponds to the prediction
from exact form factors.}
\end{figure}
\begin{figure}
\begin{centering}\psfrag{fd33}{$|f_{33}|$}
\psfrag{l}{$l$}
\psfrag{tcsa1}{$\langle\{1,0,-1\}|\Phi|\{2,0,-2\}\rangle$}
\psfrag{tcsa2}{$\langle\{1,0,-1\}|\Phi|\{3,0,-3\}\rangle$}
\psfrag{tcsa3}{$\langle\{2,0,-2\}|\Phi|\{3,0,-3\}\rangle$}\includegraphics[scale=1.2]{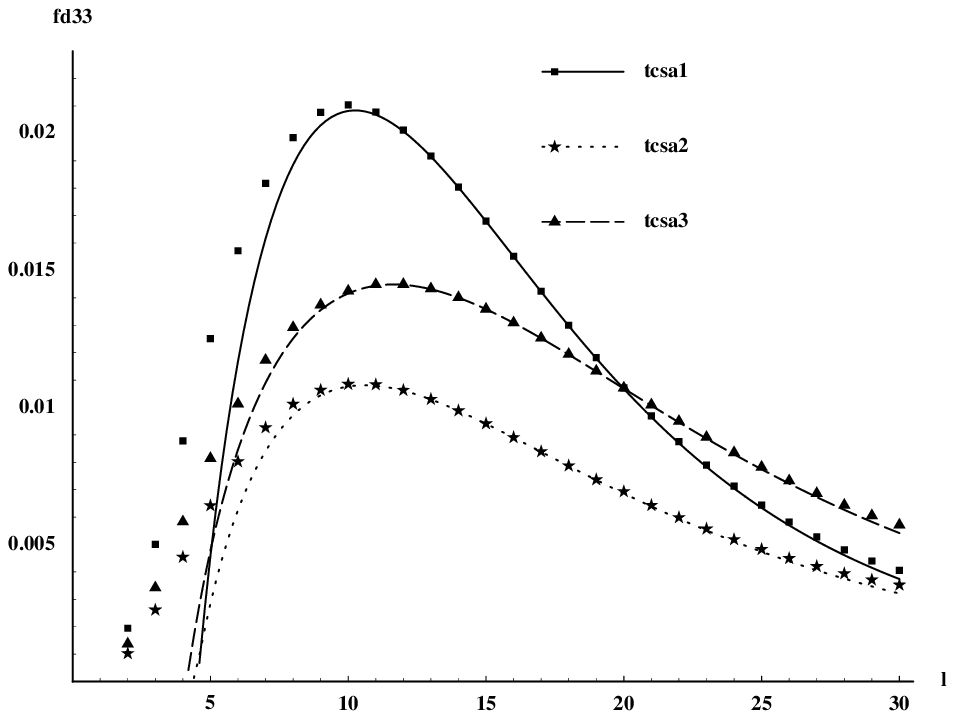}\par\end{centering}

\caption{\label{fig:fd33ly}$3$-particle--$3$-particle matrix elements in
the scaling Lee-Yang model. The discrete points correspond to the
TCSA data, while the continuous line corresponds to the prediction
from exact form factors.}
\end{figure}

The above two arguments cannot be considered as a proof; we do not
have a proper derivation of relation (\ref{eq:oddoddlyrule}) at the
moment. On the other hand, as we now show it agrees very well with
numerical data which would be impossible if there were some additional
$\varphi$ terms present; such terms, as shown in our previous work
\cite{fftcsa} would contribute corrections of order $1/l$ in terms
of the dimensionless volume parameter $l=mL$. 

\begin{figure}
\psfrag{f25}{$|f_{25}|$}\psfrag{f35}{$|f_{35}|$}
\psfrag{l}{$l$}
\psfrag{tcsa33}{$\langle\{\frac32,-\frac32\}|\Phi|\{2,1,0,-1,-2\}\rangle$}
\psfrag{tcsa53}{$\langle\{\frac52,-\frac32\}|\Phi|\{2,1,0,-1,-2\}\rangle$}
\psfrag{tcsathp}{$\langle\{3,-1,-2\}|\Phi|\{2,1,0,-1,-2\}\rangle$}\includegraphics[scale=0.8]{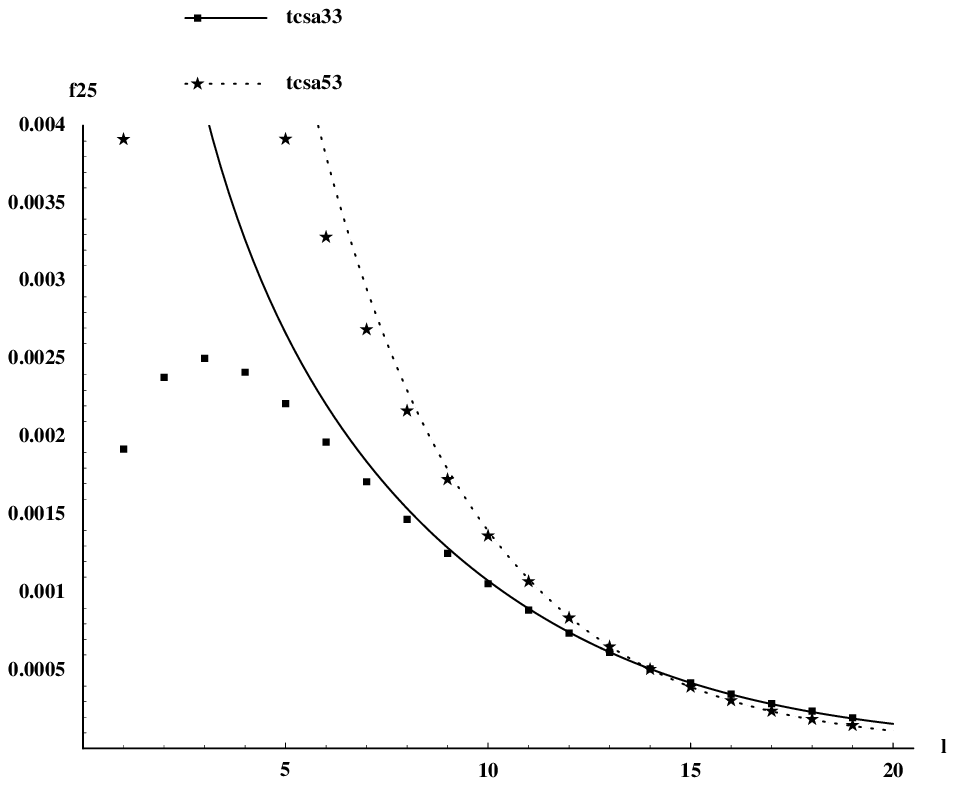}~~\includegraphics[scale=0.8]{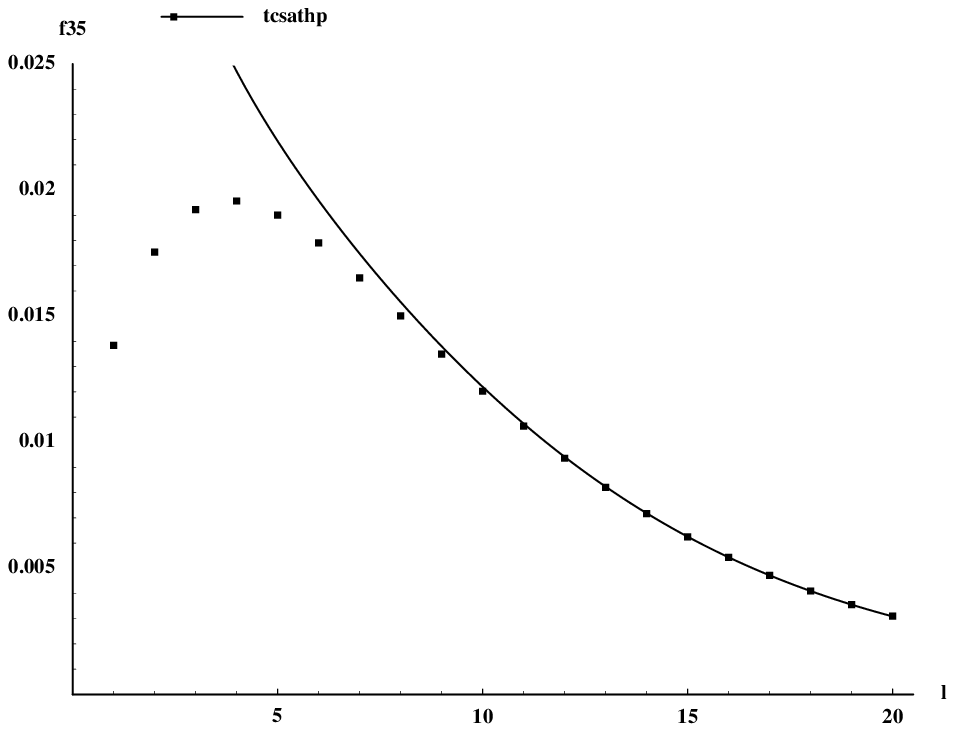}

\caption{\label{fig:5ptident} Identifying the $5$-particle state using form
factors. The discrete points correspond to the TCSA data, while the
continuous line corresponds to the prediction from exact form factors.}
\end{figure}

Data for the case of $1$-$3$ and $3$-$3$ matrix elements are shown
in figures \ref{fig:fd13ly} and \ref{fig:fd33ly}, respectively.
In order to strengthen the support for eqn. (\ref{eq:oddoddlyrule})
we must find $5$-particle states. This is not easy because they are
high up in the spectrum, and identification using the process of matching
against Bethe-Yang predictions (as described in \cite{fftcsa}) becomes
ambiguous. We could identify the first $5$-particle state by combining
the Bethe-Yang matching with predictions for matrix elements with
no disconnected pieces given by eqn. (\ref{eq:genffrelation}), as
shown in figure \ref{fig:5ptident}. Some care must be taken in choosing
the other state because many choices give matrix elements that are
too small to be measured reliably in TCSA: since vector components
and TCSA matrices are mostly of order $1$ or slightly less, getting
a result of order $10^{-4}$ or smaller involves a lot of cancellation
between a large number of individual contributions, which inevitably
leads to the result being dominated by truncation errors. Despite
these difficulties, combining Bethe-Yang level matching with form
factor evaluation we could identify the first five-particle level
up to $l=20$.

\begin{figure}
\begin{centering}\psfrag{fd13}{$|f_{35}|$}
\psfrag{l}{$l$}
\psfrag{tcsa1}{$\langle\{1,0,-1\}|\Phi|\{2,1,0,-1,-2\}\rangle$}
\psfrag{tcsa2}{$\langle\{2,0,-2\}|\Phi|\{2,1,0,-1,-2\}\rangle$}
\psfrag{tcsa3}{$\langle\{3,0,-3\}|\Phi|\{2,1,0,-1,-2\}\rangle$}\includegraphics[scale=1.2]{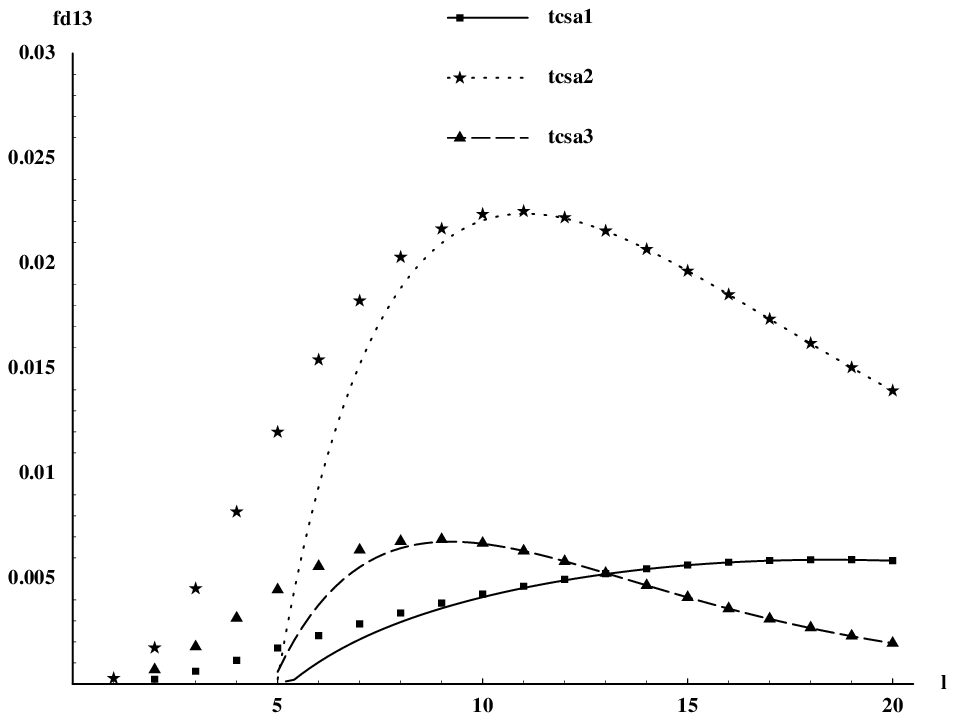}\par\end{centering}

\caption{\label{fig:fd35ly}$3$-particle--$5$-particle matrix elements in
the scaling Lee-Yang model. The discrete points correspond to the
TCSA data, while the continuous line corresponds to the prediction
from exact form factors.}
\end{figure}

The simplest matrix element involving a five-particle state and zero-momentum
disconnected pieces is the $1$-$5$ one, but the prediction of eqn.
(\ref{eq:oddoddlyrule}) turns out to be too small to be usefully
compared to TCSA. However, it is possible to find $3$-$5$ matrix
elements that are sufficiently large, and the data shown in figure
\ref{fig:fd35ly} confirm our conjecture with a relative precision
of somewhat better than $10^{-3}$ in the scaling region.

We close by noting that since the agreement is better than one part
in $10^{3}$ in the scaling region, which is typically found in the
range of volume $l\sim10\dots20$, and also this precision holds for
quite a large number of independent matrix elements, the presence
of additional $\varphi$ terms in eqn. (\ref{eq:oddoddlyrule}) can
be confidently excluded.

\subsection{Ising model in magnetic field}

\begin{figure}
\noindent \begin{centering}\psfrag{l}{$l$}
\psfrag{ff1111}{$|f_{1,111}|$}
\psfrag{a0---a0a0a0}{${}_{1}\langle\{0\}|\Psi|\{1,0,-1\}\rangle_{111}$}\includegraphics{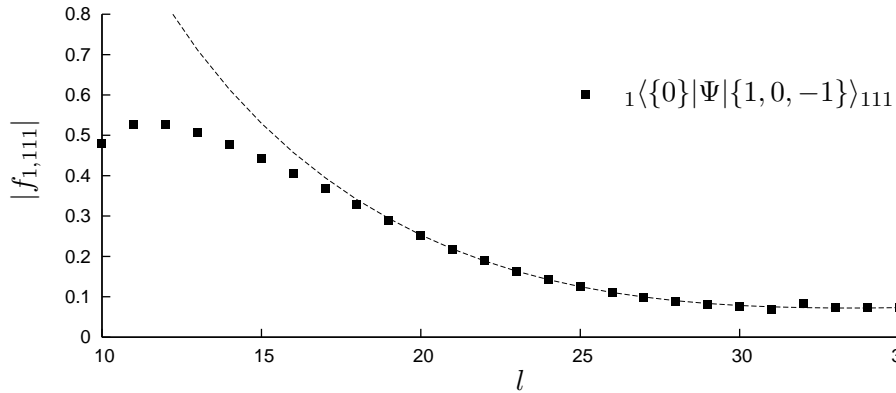}\par\end{centering}

\caption{\label{fig:fd13ising}$A_{1}-A_{1}A_{1}A_{1}$ matrix element in
Ising model with a zero-momentum particle}
\end{figure}

In figure \ref{fig:fd13ising} we show how the prediction (\ref{eq:oddoddlyrule})
describes a $1$-$3$ matrix element in the Ising model; since all
particles in this example are of species $A_{1}$, the formula carries
over without essential modifications.

\begin{figure}
\noindent \begin{centering}\psfrag{l}{$l$}
\psfrag{ff112}{$|f_{1,12}|$}
\psfrag{a0---a0b0}{${}_{1}\langle\{0\}|\Psi|\{0,0\}\rangle_{12}$}\includegraphics{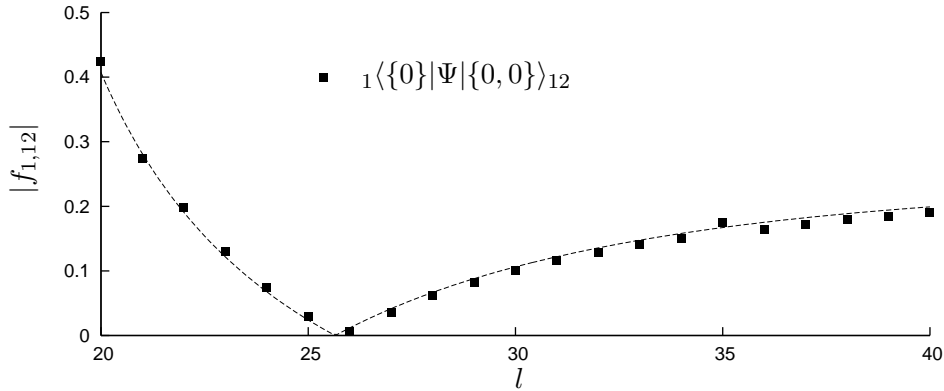}\par\end{centering}

\caption{\label{fig:fd12ising}$A_{1}-A_{1}A_{2}$ matrix element in Ising
model with zero-momentum particle}
\end{figure}

However, due to the fact that the Ising model has more than one particle
species, it is possible to have more than one stationary particles
in the same state. Our TCSA data allow us to locate one such state,
with a stationary $A_{1}$ and $A_{2}$ particle, and extending our
previous considerations we have the prediction\[
f_{1,12}={}_{1}\langle\{0\}|\Psi|\{0,0\}\rangle_{12}=\frac{1}{m_{1}L\sqrt{m_{2}L}}\left(\lim_{\epsilon\rightarrow0}F_{3}(i\pi+\epsilon,0,0)_{112}+m_{1}L\, F_{1}(0)_{2}\right)\]
where $F_{1}(0)_{2}$ is the one-particle form factor corresponding
to $A_{2}$. This is compared to TCSA data in figure \ref{fig:fd12ising}
and a convincing agreement is found.

Note that in both of figures \ref{fig:fd13ising} and \ref{fig:fd12ising}
there is a point which obviously deviates from the prediction. This
is a purely technical issue, and is due to the presence of a line
crossing close to this particular value of the volume which makes
the cutoff dependence more complicated and so slightly upsets the
extrapolation in the cutoff. We also remark that we cannot check further
matrix elements at the moment, because the appropriate form factor
solutions have not yet been computed. 

\section{Finite temperature correlators}

In this section we show how a systematical low-temperature expansion
for correlation functions can be developed using the results presented
so far. Finite temperature correlation functions have attracted quite
a lot of interest recently. Leclair and Mussardo proposed an expansion
for the one-point and two-point functions in terms of form factors
dressed by appropriate occupation number factors containing the TBA
pseudo-energy function \cite{leclairmussardo}, based on a quasi-particle
description motivated by the thermodynamic Bethe Ansatz. As discussed
in the introduction, their proposal for the two-point function was
shown to be incorrect by Saleur \cite{saleurfiniteT}; on the other
hand, he also gave a proof of the Leclair-Mussardo formula for one-point
functions based on the conjecture formulated in eqn. (\ref{eq:diaggenrulesaleur}),
provided the operator considered is the density of some local conserved
charge. Since we proved that our formula (\ref{eq:diaggenrule}) for
diagonal matrix elements is equivalent to Saleur's conjecture, our
results in section 4 can be considered as a very convincing numerical
evidence for the correctness of his argument.

Another proposal for finite-temperature one-point functions was made
by Delfino \cite{delfinofiniteT}, who attempted to express them in
terms of free-particle occupation numbers and the symmetric evaluation
of diagonal matrix elements. It was shown by Mussardo that this proposal
is not correct using a counter example where it disagreed with the
Leclair-Mussardo expansion \cite{mussardodifference}. 

Furthermore, Castro-Alvaredo and Fring also argued \cite{castrofring}
that two-point functions cannot be obtained by a simple dressing procedure
analogous to the Leclair-Mussardo expansion for one-point functions.
They argued that one needs a more drastic change in the form factor
program.

All these issues are connected to the problem of finding a proper
definition of the disconnected pieces. From the crossing relation
(\ref{eq:ffcrossing}), these are infinite for the form factors defined
in infinite volume, and subtraction of such infinities must be made
with care in order to obtain the correct finite pieces. Because of
the above difficulties there is also a development in the direction
of finite temperature form factors (for a review cf. \cite{Doyon});
with further development, this other line of thought can also give
a very useful formulation of finite temperature correlation functions.

Here we use the idea that putting the system into a finite volume
$L$ provides a regularization for the form factors, which can even
be considered physical since in the real world there are no infinite
systems%
\footnote{There is actually a little subtlety here, since we impose periodic
boundary conditions which are also nonphysical, but we make use of
the old intuition that nothing can actually depend very much on the
choice of the boundary condition if the system is very large and has
a finite correlation length (i.e. a mass gap).%
}. Our expressions for the finite volume form factors are valid up
to exponential corrections in the volume, which makes it clear that
performing the calculation in finite volume and then taking the limit
$L\rightarrow\infty$ we should recover the proper finite temperature
correlation function. Here we present the computation for the case
of the one-point function up to the first three nontrivial orders;
the calculation gets complicated for higher orders, but the recipe
is straightforward. On general theoretical grounds, it is quite clear
that our approach should also apply to the two-point function, or
indeed to any multi-point correlator, but in order to keep the exposition
short we do not go into these details here and leave them to future
investigations.

\subsection{Leclair-Mussardo series expanded}

The finite temperature expectation value of a local operator $\mathcal{O}$
is defined by\[
\langle\mathcal{O}\rangle^{R}=\frac{\text{Tr}\left(\mathrm{e}^{-RH}\mathcal{O}\right)}{\text{Tr}\left(\mathrm{e}^{-RH}\right)}\]
where $R=1/T$ is the temperature dependent extension of the Euclidean
time direction used in thermal quantum field theory and $H$ is the
Hamiltonian. To keep the exposition simple we assume that the spectrum
contains a single massive particle of mass $m$. Leclair and Mussardo
proposed the following expression for the low temperature ($T\ll m$,
or equivalently $mR\gg1$) expansion of the above one-point function:\begin{equation}
\langle\mathcal{O}\rangle^{R}=\sum_{n=0}^{\infty}\frac{1}{n!}\frac{1}{(2\pi)^{n}}\int\left[\prod_{i=1}^{n}d\theta_{i}\frac{\mathrm{e}^{-\epsilon(\theta_{i})}}{1+\mathrm{e}^{-\epsilon(\theta_{i})}}\right]F_{2n}^{c}(\theta_{1},...,\theta_{n})\label{eq:leclmuss1pt}\end{equation}
where $F_{2n}^{c}$ is the connected diagonal form factor defined
in eqn. (\ref{eq:connected}) and $\epsilon(\theta)$ is the pseudo-energy
function, which is the solution of the thermodynamic Bethe Ansatz
equation

\begin{equation}
\epsilon(\theta)=mR\cosh(\theta)-\int\frac{d\theta'}{2\pi}\varphi(\theta-\theta')\log(1+\mathrm{e}^{-\epsilon(\theta')})\label{eq:TBA}\end{equation}
The solution of this equation can be found by successive iteration,
which results in \begin{eqnarray}
\epsilon(\theta) & = & mR\cosh(\theta)-\int\frac{d\theta'}{2\pi}\varphi(\theta-\theta')\mathrm{e}^{-mR\cosh\theta'}+\frac{1}{2}\int\frac{d\theta'}{2\pi}\varphi(\theta-\theta')\mathrm{e}^{-2mR\cosh\theta'}+\nonumber \\
 & + & \int\frac{d\theta'}{2\pi}\frac{d\theta''}{2\pi}\varphi(\theta-\theta')\varphi(\theta'-\theta'')\mathrm{e}^{-mR\cosh\theta'}\mathrm{e}^{-mR\cosh\theta''}+O\left(\mathrm{e}^{-3mR}\right)\label{eq:pseudoenergy2}\end{eqnarray}
Using this expression, it is easy to derive the following expansion
from (\ref{eq:leclmuss1pt})\begin{eqnarray}
\langle\mathcal{O}\rangle^{R} & = & \langle\mathcal{O}\rangle+\int\frac{d\theta}{2\pi}F_{2}^{c}\left(\mathrm{e}^{-mR\cosh{\theta}}-\mathrm{e}^{-2mR\cosh{\theta}}\right)\nonumber \\
 &  & +\frac{1}{2}\int\frac{d\theta_{1}}{2\pi}\frac{d\theta_{2}}{2\pi}\left(F_{4}^{c}(\theta_{1},\theta_{2})+2\Phi(\theta_{1}-\theta_{2})F_{2}^{c}\right)\mathrm{e}^{-mR\cosh{\theta_{1}}}\mathrm{e}^{-mR\cosh{\theta_{2}}}\nonumber \\
 &  & +O\left(\mathrm{e}^{-3mR}\right)\label{eq:TBA2c}\end{eqnarray}
where $\langle\mathcal{O}\rangle$ denotes the zero-temperature vacuum
expectation value. The above result can also be written in terms of
the symmetric evaluation (\ref{eq:Fs_definition}) as

\begin{eqnarray}
\langle\mathcal{O}\rangle^{R} & = & \langle\mathcal{O}\rangle+\int\frac{d\theta}{2\pi}F_{2}^{s}\left(\mathrm{e}^{-mR\cosh{\theta}}-\mathrm{e}^{-2mR\cosh{\theta}}\right)+\nonumber \\
 &  & \frac{1}{2}\int\frac{d\theta_{1}}{2\pi}\frac{d\theta_{2}}{2\pi}F_{4}^{s}(\theta_{1},\theta_{2})\mathrm{e}^{-mR(\cosh{\theta_{1}}+\cosh{\theta_{2}})}+O\left(\mathrm{e}^{-3mR}\right)\label{eq:TBA2s}\end{eqnarray}
where we used relations (\ref{eq:fs2fc2}) and (\ref{cs_four}). 

For completeness we also quote Delfino's proposal:\begin{equation}
\langle\mathcal{O}\rangle_{D}^{R}=\sum_{n=0}^{\infty}\frac{1}{n!}\frac{1}{(2\pi)^{n}}\int\left[\prod_{i=1}^{n}d\theta_{i}\frac{\mathrm{e}^{-mR\cosh\theta_{i}}}{1+\mathrm{e}^{-mR\cosh\theta_{i}}}\right]F_{2n}^{s}(\theta_{1},...,\theta_{n})\label{eq:delfino1pt}\end{equation}
which gives the following result when expanded to second order: \begin{eqnarray}
\langle\mathcal{O}\rangle_{D}^{R} & = & \langle\mathcal{O}\rangle+\int\frac{d\theta}{2\pi}F_{2}^{s}\left(\mathrm{e}^{-mR\cosh{\theta}}-\mathrm{e}^{-2mR\cosh{\theta}}\right)+\nonumber \\
 &  & \frac{1}{2}\int\frac{d\theta_{1}}{2\pi}\frac{d\theta_{2}}{2\pi}F_{4}^{s}(\theta_{1},\theta_{2})\mathrm{e}^{-mR(\cosh{\theta_{1}}+\cosh{\theta_{2}})}+O\left(\mathrm{e}^{-3mR}\right)\label{eq:delfino2s}\end{eqnarray}
Note that the two formulae coincide with each other to this order,
which was already noted in \cite{delfinofiniteT}. However, this is
not the case in the next order. Obtaining the third order correction
from the Leclair-Mussardo expansion is a somewhat lengthy, but elementary
computation, which results in\begin{eqnarray}
 &  & \frac{1}{6}\int\frac{d\theta_{1}}{2\pi}\frac{d\theta_{2}}{2\pi}\frac{d\theta_{3}}{2\pi}F_{6}^{s}(\theta_{1},\theta_{2},\theta_{3})\mathrm{e}^{-mR(\cosh\theta_{1}+\cosh\theta_{2}+\cosh\theta_{3})}\nonumber \\
 &  & -\int\frac{d\theta_{1}}{2\pi}\frac{d\theta_{2}}{2\pi}F_{4}^{s}(\theta_{1},\theta_{2})\mathrm{e}^{-mR(\cosh\theta_{1}+2\cosh\theta_{2})}+\int\frac{d\theta_{1}}{2\pi}F_{2}^{s}\mathrm{e}^{-3mR\cosh\theta_{1}}\nonumber \\
 &  & -\frac{1}{2}\int\frac{d\theta_{1}}{2\pi}\frac{d\theta_{2}}{2\pi}F_{2}^{s}\varphi(\theta_{1}-\theta_{2})\mathrm{e}^{-mR(\cosh\theta_{1}+2\cosh\theta_{2})}\label{eq:lm3order}\end{eqnarray}
where we used eqns. (\ref{eq:fs2fc2}, \ref{cs_four}, \ref{eq:f6sa})
to express the result in terms of the symmetric evaluation. On the
other hand, expanding (\ref{eq:delfino1pt}) results in \begin{eqnarray}
 &  & \frac{1}{6}\int\frac{d\theta_{1}}{2\pi}\frac{d\theta_{2}}{2\pi}\frac{d\theta_{3}}{2\pi}F_{6}^{s}(\theta_{1},\theta_{2},\theta_{3})\mathrm{e}^{-mR(\cosh\theta_{1}+\cosh\theta_{2}+\cosh\theta_{3})}\nonumber \\
 &  & -\int\frac{d\theta_{1}}{2\pi}\frac{d\theta_{2}}{2\pi}F_{4}^{s}(\theta_{1},\theta_{2})\mathrm{e}^{-mR(\cosh\theta_{1}+2\cosh\theta_{2})}+\int\frac{d\theta_{1}}{2\pi}F_{2}^{s}\mathrm{e}^{-3mR\cosh\theta_{1}}\label{eq:delfino3order}\end{eqnarray}
It can be seen that the two proposals differ at this order (the last
term of (\ref{eq:lm3order}) is missing from (\ref{eq:delfino3order})),
which was already noted by Mussardo using a toy model in \cite{mussardodifference},
but our computation here is model independent and shows the general
form of the discrepancy. We also need the third order correction explicitly
so that we can compare it to the result of the computation performed
in the next section.

\subsection{Low-temperature expansion for one-point functions}

We now evaluate the finite temperature expectations value in a finite,
but large volume $L$:

\begin{equation}
\langle\mathcal{O}\rangle_{L}^{R}=\frac{\text{Tr}_{L}\left(\mathrm{e}^{-RH_{L}}\mathcal{O}\right)}{\text{Tr}_{L}\left(\mathrm{e}^{-RH_{L}}\right)}\label{onepointRL}\end{equation}
where $H_{L}$ is the finite volume Hamiltonian, and $\mathrm{Tr}_{L}$
means that the trace is now taken over the finite volume Hilbert space.
For later convenience we introduce a new notation:\[
|\theta_{1},\dots,\theta_{n}\rangle_{L}=|\{ I_{1},\dots,I_{n}\}\rangle_{L}\]
where $\theta_{1},\dots,\theta_{n}$ solve the Bethe-Yang equations
for $n$ particles with quantum numbers $I_{1},\dots,I_{n}$ at the
given volume $L$. We can develop the low temperature expansion of
(\ref{onepointRL}) in powers of $\mathrm{e}^{-mR}$ using

\begin{eqnarray}
\text{Tr}_{L}\left(\mathrm{e}^{-RH_{L}}\mathcal{O}\right) & = & \langle\mathcal{O}\rangle_{L}+\sum_{\theta^{(1)}}\mathrm{e}^{-mR\cosh\theta^{(1)}}\langle\theta^{(1)}|\mathcal{O}|\theta^{(1)}\rangle_{L}\nonumber \\
 &  & +\frac{1}{2}\sum_{\theta_{1}^{(2)},\theta_{2}^{(2)}}{}^{'}\mathrm{e}^{-mR(\cosh\theta_{1}^{(2)}+\cosh\theta_{2}^{(2)})}\langle\theta_{1}^{(2)},\theta_{2}^{(2)}|\mathcal{O}|\theta_{1}^{(2)},\theta_{2}^{(2)}\rangle_{L}+\nonumber \\
 &  & +\frac{1}{6}\sum_{\theta_{1}^{(3)},\theta_{2}^{(3)},\theta_{3}^{(3)}}{}^{'}\mathrm{e}^{-mR(\cosh\theta_{1}^{(3)}+\cosh\theta_{2}^{(3)}+\cosh\theta_{3}^{(3)})}\langle\theta_{1}^{(3)},\theta_{2}^{(3)},\theta_{3}^{(3)}|\mathcal{O}|\theta_{1}^{(3)},\theta_{2}^{(3)},\theta_{3}^{(3)}\rangle_{L}\nonumber \\
 &  & +O(\mathrm{e}^{-4mR})\label{eq:nomexp}\end{eqnarray}
and\begin{eqnarray}
\text{Tr}_{L}\left(\mathrm{e}^{-RH_{L}}\right) & = & 1+\sum_{\theta^{(1)}}\mathrm{e}^{-mR\cosh(\theta^{(1)})}+\frac{1}{2}\sum_{\theta_{1}^{(2)},\theta_{2}^{(2)}}{}^{'}\mathrm{e}^{-mR(\cosh(\theta_{1}^{(2)})+\cosh(\theta_{2}^{(2)}))}\nonumber \\
 &  & +\frac{1}{6}\sum_{\theta_{1}^{(3)},\theta_{2}^{(3)},\theta_{3}^{(3)}}{}^{'}\mathrm{e}^{-mR(\cosh\theta_{1}^{(3)}+\cosh\theta_{2}^{(3)}+\cosh\theta_{3}^{(3)})}+O(\mathrm{e}^{-4mR})\label{eq:Zexp}\end{eqnarray}
The denominator of (\ref{onepointRL}) can then be easily expanded:\begin{eqnarray}
\frac{1}{\text{Tr}_{L}\left(\mathrm{e}^{-RH_{L}}\right)} & = & 1-\sum_{\theta^{(1)}}\mathrm{e}^{-mR\cosh\theta^{(1)}}+\left(\sum_{\theta^{(1)}}\mathrm{e}^{-mR\cosh\theta^{(1)}}\right)^{2}-\frac{1}{2}\sum_{\theta_{1}^{(2)},\theta_{2}^{(2)}}{}^{'}\mathrm{e}^{-mR(\cosh\theta_{1}^{(2)}+\cosh\theta_{2}^{(2)})}\nonumber \\
 &  & -\left(\sum_{\theta^{(1)}}\mathrm{e}^{-mR\cosh\theta^{(1)}}\right)^{3}+\left(\sum_{\theta^{(1)}}\mathrm{e}^{-mR\cosh\theta^{(1)}}\right)\sum_{\theta_{1}^{(2)},\theta_{2}^{(2)}}{}^{'}\mathrm{e}^{-mR(\cosh\theta_{1}^{(2)}+\cosh\theta_{2}^{(2)})}\nonumber \\
 &  & -\frac{1}{6}\sum_{\theta_{1}^{(3)},\theta_{2}^{(3)},\theta_{3}^{(3)}}{}^{'}\mathrm{e}^{-mR(\cosh\theta_{1}^{(3)}+\cosh\theta_{2}^{(3)}+\cosh\theta_{3}^{(3)})}+O(\mathrm{e}^{-4mR})\label{eq:Zinvexp}\end{eqnarray}
The primes in the multi-particle sums serve as a reminder that there
exist only states for which all quantum numbers are distinct. Since
we assumed that there is a single particle species, this means that
terms in which any two of the rapidities coincide are excluded. All
$n$-particle terms in (\ref{eq:nomexp}) and (\ref{eq:Zexp}) have
a $1/n!$ prefactor which takes into account that different ordering
of the same rapidities give the same state; as the expansion contains
only diagonal matrix elements, phases resulting from reordering the
particles cancel. The upper indices of the rapidity variables indicate
the number of particles in the original finite volume states; this
is going to be handy when replacing the discrete sums with integrals
since it keeps track of which multi-particle state density is relevant. 

We also need an extension of the finite volume matrix elements to
rapidities that are not necessarily solutions of the appropriate Bethe-Yang
equations. The required analytic continuation is simply given by eqn.
(\ref{eq:diaggenrule})

\begin{equation}
\langle\theta_{1},\dots,\theta_{n}|\mathcal{O}|\theta_{1},\dots,\theta_{n}\rangle_{L}=\frac{1}{\rho_{n}(\theta_{1},\dots,\theta_{n})_{L}}\,\sum_{A\subset\{1,2,\dots n\}}F_{2|A|}^{s}(\{\theta_{i}\}_{i\in A})\rho_{n-|A|}(\{\theta_{i}\}_{i\notin A})_{L}+O(\mathrm{e}^{-\mu L})\label{eq:Fsextended}\end{equation}
where we made explicit the volume dependence of the $n$-particle
density factors. The last term serves as a reminder that this prescription
only defines the form factor to all orders in $1/L$ (i.e. up to residual
finite size corrections), but this is sufficient to perform the computations
in the sequel.

Using the leading behaviour of the $n$-particle state density, contributions
from the $n$-particle sector scale as $L^{n}$, and for the series
expansions (\ref{eq:nomexp}), (\ref{eq:Zexp}) and (\ref{eq:Zinvexp})
it is necessary that $mL\ll\mathrm{e}^{mR}$. However if $mR$ is
big enough there remains a large interval \[
1\ll mL\ll\mathrm{e}^{mR}\]
where the expansions are expected to be valid. After substituting
these expansions into (\ref{onepointRL}) we will find order by order
that the leading term of the net result is $O(L^{0})$, and the corrections
scale as negative powers of $L$. Therefore in (\ref{onepointRL})
we can continue analytically to large $L$ and take the $L\rightarrow\infty$
limit.

\subsubsection{Corrections of order $\mathrm{e}^{-mR}$}

Substituting the appropriate terms from (\ref{eq:Zinvexp}) and (\ref{eq:nomexp})
into (\ref{onepointRL}) gives the result\[
\langle\mathcal{O}\rangle_{L}^{R}=\langle\mathcal{O}\rangle_{L}+\sum_{\theta^{(1)}}\mathrm{e}^{-mR\cosh\theta^{(1)}}\left(\langle\theta^{(1)}|\mathcal{O}|\theta^{(1)}\rangle_{L}-\langle\mathcal{O}\rangle_{L}\right)+O(\mathrm{e}^{-2mR})\]
Taking the $L\to\infty$ limit one can replace the summation with
an integral over the states in the rapidity space: \[
\sum_{i}\to\int\frac{d\theta}{2\pi}\rho_{1}(\theta)\]
and using (\ref{eq:d1formula}) we can write

\begin{equation}
\rho_{1}(\theta)\left(\langle\theta|\mathcal{O}|\theta\rangle_{L}-\langle\mathcal{O}\rangle_{L}\right)=F_{2}^{s}+O(\mathrm{e}^{-\mu L})\label{1pbol1}\end{equation}
so we obtain\[
\langle\mathcal{O}\rangle^{R}=\langle\mathcal{O}\rangle+\int\frac{d\theta}{2\pi}F_{2}^{s}\mathrm{e}^{-mR\cosh{\theta}}+O(\mathrm{e}^{-2mR})\]
which coincides with eqn. (\ref{eq:TBA2s}) to this order.

\subsubsection{Corrections of order $e^{-2mR}$}

Substituting again the appropriate terms from (\ref{eq:Zinvexp})
and (\ref{eq:nomexp}) into (\ref{onepointRL}) gives the result\begin{eqnarray*}
\langle\mathcal{O}\rangle_{L}^{R} & = & \langle\mathcal{O}\rangle_{L}+\sum_{\theta^{(1)}}\mathrm{e}^{-mR\cosh\theta^{(1)}}\left(\langle\theta^{(1)}|\mathcal{O}|\theta^{(1)}\rangle_{L}-\langle\mathcal{O}\rangle_{L}\right)\\
 &  & -\left(\sum_{\theta_{1}^{(1)}}\mathrm{e}^{-mR\cosh\theta_{1}^{(1)}}\right)\left(\sum_{\theta_{2}^{(1)}}\mathrm{e}^{-mR\cosh\theta_{2}^{(1)}}\left(\langle\theta_{2}^{(1)}|\mathcal{O}|\theta_{2}^{(1)}\rangle_{L}-\langle\mathcal{O}\rangle_{L}\right)\right)\\
 &  & +\frac{1}{2}\sum_{\theta_{1}^{(2)},\theta_{2}^{(2)}}{}^{'}\mathrm{e}^{-mR(\cosh\theta_{1}^{(2)}+\cosh\theta_{2}^{(2)})}\left(\langle\theta_{1}^{(2)},\theta_{2}^{(2)}|\mathcal{O}|\theta_{1}^{(2)},\theta_{2}^{(2)}\rangle_{L}-\langle\mathcal{O}\rangle_{L}\right)+O(\mathrm{e}^{-3mR})\end{eqnarray*}
The $O(\mathrm{e}^{-2mR})$ terms can be rearranged as follows. We
add and subtract a term to remove the constraint from the two-particle
sum:\begin{eqnarray*}
 &  & +\frac{1}{2}\sum_{\theta_{1}^{(2)},\theta_{2}^{(2)}}\mathrm{e}^{-mR(\cosh\theta_{1}^{(2)}+\cosh\theta_{2}^{(2)})}\left(\langle\theta_{1}^{(2)},\theta_{2}^{(2)}|\mathcal{O}|\theta_{1}^{(2)},\theta_{2}^{(2)}\rangle_{L}-\langle\mathcal{O}\rangle_{L}\right)\\
 &  & -\frac{1}{2}\sum_{\theta_{1}^{(2)}=\theta_{2}^{(2)}}\mathrm{e}^{-2mR\cosh\theta_{1}^{(2)}}\left(\langle\theta_{1}^{(2)},\theta_{1}^{(2)}|\mathcal{O}|\theta_{1}^{(2)},\theta_{1}^{(2)}\rangle_{L}-\langle\mathcal{O}\rangle_{L}\right)\\
 &  & -\frac{1}{2}\sum_{\theta_{1}^{(1)}}\sum_{\theta_{2}^{(1)}}\mathrm{e}^{-mR(\cosh\theta_{1}^{(1)}+\cosh\theta_{2}^{(1)})}\left(\langle\theta_{1}^{(1)}|\mathcal{O}|\theta_{1}^{(1)}\rangle_{L}+\langle\theta_{2}^{(1)}|\mathcal{O}|\theta_{2}^{(1)}\rangle_{L}-2\langle\mathcal{O}\rangle_{L}\right)\end{eqnarray*}
The $\theta_{1}^{(2)}=\theta_{2}^{(2)}$ terms correspond to insertion
of some spurious two-particle states with equal Bethe quantum numbers
for the two particles ($I_{1}=I_{2}$). The two-particle Bethe-Yang
equations in this case degenerates to the one-particle case (as discussed
before, the matrix elements can be defined for these {}``states''
without any problems since we have the analytic formula (\ref{eq:Fsextended})
valid to any order in $1/L$). This also means that the density relevant
to the diagonal two-particle sum is $\rho_{1}$ and so for large $L$
we can substitute the sums with the following integrals \[
\sum_{\theta_{1,2}^{(1)}}\rightarrow\int\frac{d\theta_{1,2}}{2\pi}\rho_{1}(\theta_{1,2})\quad,\quad\sum_{\theta_{1}^{(2)}=\theta_{2}^{(2)}}\rightarrow\int\frac{d\theta}{2\pi}\rho_{1}(\theta)\quad,\quad\sum_{\theta_{1}^{(2)},\theta_{2}^{(2)}}\rightarrow\int\frac{d\theta_{1}}{2\pi}\frac{d\theta_{2}}{2\pi}\rho_{2}(\theta_{1.},\theta_{2})\]
Let us express the finite volume matrix elements in terms of form
factors using (\ref{eq:d1formula}) and (\ref{eq:d2formula}):\begin{eqnarray*}
 &  & \rho_{2}(\theta_{1},\theta_{2})\left(\langle\theta_{1}^{(2)},\theta_{2}^{(2)}|\mathcal{O}|\theta_{1}^{(2)},\theta_{2}^{(2)}\rangle_{L}-\langle\mathcal{O}\rangle_{L}\right)\\
 &  & -\rho_{1}\left(\theta_{1}\right)\rho_{1}\left(\theta_{2}\right)\left(\langle\theta_{1}|\mathcal{O}|\theta_{1}\rangle_{L}+\langle\theta_{2}|\mathcal{O}|\theta_{2}\rangle_{L}-2\langle\mathcal{O}\rangle_{L}\right)=F_{4}^{s}(\theta_{1},\theta_{2})+O(\mathrm{e}^{-\mu L})\end{eqnarray*}
Combining the above relation with (\ref{1pbol1}), we also have\[
\langle\theta,\theta|\mathcal{O}|\theta,\theta\rangle_{L}-\langle\mathcal{O}\rangle_{L}=\frac{2\rho_{1}\left(\theta\right)}{\rho_{2}(\theta,\theta)}F_{2}^{s}+O(\mathrm{e}^{-\mu L})\]
where we used that $F_{4}^{s}(\theta,\theta)=0$, which is just the
exclusion property mention after eqn. (\ref{eq:Fs_definition}). Note
that\[
\frac{\rho_{1}(\theta)^{2}}{\rho_{2}(\theta,\theta)}=1+O(L^{-1})\]
and therefore in the limit $L\rightarrow\infty$ we obtain\[
-\int\frac{d\theta}{2\pi}\mathrm{e}^{-2mR\cosh\theta}F_{2}^{s}+\frac{1}{2}\int\frac{d\theta_{1}}{2\pi}\frac{d\theta_{2}}{2\pi}F_{4}^{s}(\theta_{1},\theta_{2})\mathrm{e}^{-mR(\cosh\theta_{1}+\cosh\theta_{2})}\]
which is equal to the relevant contributions in the Leclair-Mussardo
expansion (\ref{eq:TBA2s}).

\subsubsection{Corrections of order $e^{-3mR}$}

This calculation is rather long, and so it is relegated to the appendix.
The net result is\begin{eqnarray}
 &  & \frac{1}{6}\int\frac{d\theta_{1}}{2\pi}\frac{d\theta_{2}}{2\pi}\frac{d\theta_{3}}{2\pi}F_{6}^{s}(\theta_{1},\theta_{2},\theta_{3})\mathrm{e}^{-mR(\cosh\theta_{1}+\cosh\theta_{2}+\cosh\theta_{3})}\nonumber \\
 &  & -\int\frac{d\theta_{1}}{2\pi}\frac{d\theta_{2}}{2\pi}F_{4}^{s}(\theta_{1},\theta_{2})\mathrm{e}^{-mR(\cosh\theta_{1}+2\cosh\theta_{2})}+\int\frac{d\theta_{1}}{2\pi}F_{2}^{s}\mathrm{e}^{-3mR\cosh\theta_{1}}\nonumber \\
 &  & -\frac{1}{2}\int\frac{d\theta_{1}}{2\pi}\frac{d\theta_{2}}{2\pi}F_{2}^{s}\varphi(\theta_{1}-\theta_{2})\mathrm{e}^{-mR(\cosh\theta_{1}+2\cosh\theta_{2})}\label{eq:our3order}\end{eqnarray}
which agrees exactly with eqn. (\ref{eq:lm3order}).

\subsection{Remarks}

There are a few remarks which we wish to make. First, we see that
the proposals by Leclair and Mussardo and by Delfino differ at the
order $\mathrm{e}^{-3mR}$. The reason for this difference can be
understood in the formalism developed here. Namely, the expansions
(\ref{eq:nomexp}) and (\ref{eq:Zinvexp}) both contain positive powers
of $L$. On physical grounds, they are expected to cancel completely
order by order in the $\mathrm{e}^{-mR}$ expansion. However, the
state densities $\rho$ depend on the interaction as well. This dependence
is of order $L^{-1}$, and it actually characterizes the ambiguity
in the definition of the diagonal matrix element resulting from the
resolution of the singularity (see eqn. (\ref{mostgeneral})). Naively
it drops out in the $L\rightarrow\infty$ limit, but actually some
of these terms is multiplied by a positive $L$ power from (\ref{eq:Zinvexp}).
In our derivation we evaluated every relevant contribution to all
orders in $1/L$ (i.e. we only neglected residual finite size corrections).
As a result, we could take the limit $L\rightarrow\infty$ properly
and get the correct finite part of the resulting expression.

Taking this line of thought further, note that the leading term of
every multi-particle density (whether it is degenerate in the sense
defined in the appendix, or not) is always a product of $E_{i}L$
factors where $i$ runs over the number of particles and $E_{i}$
is their energy. Therefore density terms whose leading behaviour is
$L^{0}$ do not contribute explicit $\varphi$ factors. As far as
there are only contributions of this type, the expansion of the one-point
function, when written in terms of $F^{s}$ is just the same as in
a free field theory. Indeed in the free field limit the Leclair-Mussardo
expansion and the Delfino proposal are identical, since the pseudo-energy
function is just $\epsilon(\theta)=mR\cosh\theta$ and $F_{2n}^{c}\equiv F_{2n}^{s}$
(more generally, due to the absence of kinematical singularities the
$\epsilon_{i}\rightarrow0$ limit of (\ref{mostgeneral}) is independent
of the direction). 

To have terms that depend explicitly on the interaction we need density
contributions that naively scale as a positive power of $L$. When
combining all such terms at a given order, the leading term must drop
out, and the final result can only have a behaviour $L^{0}$ at large
$L$. It is clear from our calculation detail above and in the appendix
that the first order at which such an anomalous contribution arises
is that of $\mathrm{e}^{-3mR}$. Up to that order every individual
term is finite as $L\rightarrow\infty$. However, at third order there
appear some {}``anomalous'' density terms, namely those collected
in (\ref{eq:anomdens}), which individually grow linearly in $L$.
As required by general principles, the linear contribution cancels
between them and so the $L\rightarrow\infty$ limit is well-defined.
However, the subleading terms always contain dependence on $\varphi$,
and indeed they all vanish for a free theory (when $\varphi=0$),
therefore it is only such terms that can contribute explicit $\varphi$
dependence in the expansion. As a result, there remains an {}``anomalous''
term which is just ($-1$ times) the derivative of the phase shift,
and leads to the correction (\ref{eq:dintres2}), which is exactly
the term absent in Delfino's expression.

Strictly speaking, the above discussion is only valid if the expansion
is written in terms of the symmetric evaluation $F_{2n}^{s}$ ; rewriting
it in terms of the connected form factors $F_{2n}^{c}$ obviously
introduces further $\varphi$ dependence. As shown in the above argument,
the real difference between the free and the interacting case can
be properly observed when the expansion is written in terms of $F_{2n}^{s}$,
therefore it seems a more natural choice than using the connected
form factors, as the behaviour specific to interacting theories can
be seen much more clearly. 

Another important point is that our results give an independent support
for the Leclair-Mussardo expansion. It is known that it coincides
precisely with the exact TBA result for the trace of the energy-momentum
tensor \cite{leclairmussardo}, and Saleur presented an argument for
its validity when the operator considered is the density of a local
conserved charge \cite{saleurfiniteT}. These arguments work to all
orders, but only for a restricted set of local operators. On the other
hand, our calculation above is model independent, and although we
only worked it out to order $\mathrm{e}^{-3mR}$, we expect that it
coincides with the Leclair-Mussardo expansion to all orders. For a
complete proof we need a better understanding of its structure, which
is out of the scope of the present work.

Furthermore, our method has a straightforward extension to higher
point correlation functions. For example, a two-point correlation
function \[
\langle\mathcal{O}_{1}(x)\mathcal{O}_{2}(0)\rangle_{L}^{R}=\frac{\text{Tr}_{L}\left(\mathrm{e}^{-RH_{L}}\mathcal{O}_{1}(x)\mathcal{O}_{2}(0)\right)}{\text{Tr}_{L}\left(\mathrm{e}^{-RH_{L}}\right)}\]
can be expanded inserting two complete sets of states\begin{equation}
\text{Tr}_{L}\left(\mathrm{e}^{-RH_{L}}\mathcal{O}_{1}(x)\mathcal{O}_{2}(0)\right)=\sum_{m,n}\mathrm{e}^{-RE_{n}(L)}\langle n|\mathcal{O}(x)|m\rangle_{L}\langle m|\mathcal{O}(0)|n\rangle_{L}\label{eq:2ptexp}\end{equation}
Since we now have a complete description of finite volume matrix elements
to all orders in $1/L$, the above expression can be evaluated along
the lines presented in subsection 7.2, provided that the intermediate
state sums are properly truncated. We leave the explicit evaluation
of expansion (\ref{eq:2ptexp}) to further investigations.

Finally note that besides giving a systematic expansion in powers
of $\mathrm{e}^{-mR}$, our method also gives the $L$ dependence
to all orders in $1/L$ (i.e. up to residual finite size effects),
therefore it can also be used to study finite size corrections of
correlators in the low temperature regime.

\section{Conclusions}

In this work we completed the description of finite volume matrix
elements of local operators by considering those with disconnected
pieces. There are two types of such matrix elements, namely (1) diagonal
ones and (2) ones involving parity-invariant zero-spin states with
zero-momentum particles. Our description is valid to any order in
$1/L$ i.e. up to residual finite size corrections decaying exponentially
with the volume $L$. The precise statements were formulated in subsection
2.3 and we then gave extensive numerical evidence for them. We also
formulated and proved a general theorem relating the different possible
evaluations of diagonal matrix elements, and showed that our results
coincide with the proposal made by Saleur \cite{saleurfiniteT}.

We then showed how to perform an expansion for finite temperature
correlation functions, using the fact that finite volume acts as a
regulator for the otherwise infinite disconnected pieces. The case
we considered explicitly was that of one-point functions at finite
temperature. We evaluated the first few orders in the low temperature
expansion and showed that they coincide with the result conjectured
by Leclair and Mussardo \cite{leclairmussardo}, but are different
from Delfino's proposal \cite{delfinofiniteT} at third order. Some
important aspects of this expansion were already discussed in subsection
7.3, which we do not repeat here.

There is a number of interesting issues remaining. Our approach gives
the finite volume form factors up to residual finite size effects,
but combined with truncated conformal space one can achieve a precision
of order $10^{-4}$ in the scaling Lee-Yang model, and $10^{-3}$
in the Ising model with magnetic field. It would be interesting to
see how these results can be related to other approaches to finite
volume form factors (see \cite{finitevolFF}) and whether the picture
can be completed to give some sort of exact description in the case
of integrable field theories. It also seems worthwhile to formulate
a higher dimensional generalization of these results extending the
approach of Lellouch and Lüscher \cite{lellouch}, which is expected
to be relevant for lattice field theory.

Another open issue is to give a more concise formulation of the finite
temperature expansion discussed in section 7 that would make possible
a partial resummation to recover the Leclair-Mussardo expression (\ref{eq:leclmuss1pt})
which involves dressed form factors. 

It is even more interesting to write down the expansion for two-point
correlators following the ideas outlined in subsection 7.3; a better
method of organizing the contributions could be of great help here
as well. Results for the two-point function can be compared e.g. to
evaluation of correlation functions from truncated conformal space,
and can also be used in further development of the finite temperature
form factor program \cite{Doyon}.

\subsection*{Acknowledgments}

We wish to thank Z. Bajnok and L. Palla for useful discussions. This
research was partially supported by the Hungarian research funds OTKA
T043582, K60040 and TS044839. GT was also supported by a Bolyai J\'anos
research scholarship.

\makeatletter 
\renewcommand{\theequation}{\hbox{\normalsize\Alph{section}.\arabic{equation}}} 
\@addtoreset{equation}{section} 
\renewcommand{\thefigure}{\hbox{\normalsize\Alph{section}.\arabic{figure}}} 
\@addtoreset{figure}{section} 
\renewcommand{\thetable}{\hbox{\normalsize\Alph{section}.\arabic{table}}} 
\@addtoreset{table}{section} 
\makeatother 

\appendix

\section{$\mathrm{e}^{-3mR}$ corrections to the finite temperature one-point
function}

In order to shorten the presentation, we introduce some further convenient
notations:\begin{eqnarray*}
 &  & E_{i}=m\cosh\theta_{i}\\
 &  & \langle\theta_{1},\dots,\theta_{n}|\mathcal{O}|\theta_{1},\dots,\theta_{n}\rangle_{L}=\langle1\dots n|\mathcal{O}|1\dots n\rangle_{L}\\
 &  & \rho_{n}(\theta_{1},\dots,\theta_{n})=\rho(1\dots n)\end{eqnarray*}
Summations will be shortened to\begin{eqnarray*}
\sum_{\theta_{1}\dots\theta_{n}} & \rightarrow & \sum_{1\dots n}\\
\sum_{\theta_{1}\dots\theta_{n}}{}^{'} & \rightarrow & \sum_{1\dots n}{}^{'}\end{eqnarray*}
Given these notations, we now multiply (\ref{eq:nomexp}) with (\ref{eq:Zinvexp})
and collect the third order correction terms:\begin{eqnarray*}
 &  & \frac{1}{6}\sum_{123}{}^{'}\mathrm{e}^{-R(E_{1}+E_{2}+E_{3})}\left(\langle123|\mathcal{O}|123\rangle_{L}-\langle\mathcal{O}\rangle_{L}\right)\\
 & - & \left(\sum_{1}\mathrm{e}^{-RE_{1}}\right)\frac{1}{2}\sum_{23}{}^{'}\mathrm{e}^{-R(E_{2}+E_{3})}\left(\langle23|\mathcal{O}|23\rangle_{L}-\langle\mathcal{O}\rangle_{L}\right)\\
 & + & \left\{ \left(\sum_{1}\mathrm{e}^{-RE_{1}}\right)\left(\sum_{2}\mathrm{e}^{-RE_{2}}\right)-\frac{1}{2}\sum_{12}{}^{'}\mathrm{e}^{-R(E_{1}+E_{2})}\right\} \left(\sum_{3}\mathrm{e}^{-RE_{3}}\right)\left(\langle3|\mathcal{O}|3\rangle_{L}-\langle\mathcal{O}\rangle_{L}\right)\end{eqnarray*}
To keep trace of the state densities, we avoid combining rapidity
sums. Now we replace the constrained summations by free sums with
the diagonal contributions subtracted:\begin{eqnarray*}
\sum_{12}{}^{'} & = & \sum_{12}-\sum_{1=2}\\
\sum_{123}{}^{'} & = & \sum_{123}-\left(\sum_{1=2,3}+\sum_{2=3,1}+\sum_{1=3,2}\right)+2\sum_{1=2=3}\end{eqnarray*}
where the diagonal contributions are labeled to show which diagonal
it sums over, but otherwise the given sum is free, e.g.\[
\sum_{1=2,3}\]
shows a summation over all triplets $\theta_{1}^{(3)},\theta_{2}^{(3)},\theta_{3}^{(3)}$
where $\theta_{1}^{(3)}=\theta_{2}^{(3)}$ and $\theta_{3}^{(3)}$
runs free (it can also be equal with the other two). We also make
use of the notation\[
F(12\dots n)=F_{2n}^{s}(\theta_{1},\dots,\theta_{n})\]
so the necessary matrix elements can be written in the form\begin{eqnarray}
\rho(123)\left(\langle123|\mathcal{O}|123\rangle_{L}-\langle\mathcal{O}\rangle_{L}\right) & = & F(123)+\rho(1)F(23)+\dots+\rho(12)F(3)+\dots\nonumber \\
\rho(122)\left(\langle122|\mathcal{O}|122\rangle_{L}-\langle\mathcal{O}\rangle_{L}\right) & = & 2\rho(2)F(12)+2\rho(12)F(3)+\rho(22)F(1)\nonumber \\
\rho(111)\left(\langle111|\mathcal{O}|111\rangle_{L}-\langle\mathcal{O}\rangle_{L}\right) & = & 3\rho(111)F(1)\nonumber \\
\rho(12)\left(\langle12|\mathcal{O}|12\rangle_{L}-\langle\mathcal{O}\rangle_{L}\right) & = & F(12)+\rho(1)F(2)+\rho(2)F(1)\nonumber \\
\rho(11)\left(\langle11|\mathcal{O}|11\rangle_{L}-\langle\mathcal{O}\rangle_{L}\right) & = & 2\rho(1)F(1)\nonumber \\
\rho(1)\left(\langle1|\mathcal{O}|1\rangle_{L}-\langle\mathcal{O}\rangle_{L}\right) & = & F(1)\label{eq:matelms}\end{eqnarray}
where we used that $F$ and $\rho$ are entirely symmetric in all
their arguments, and the ellipsis in the the first line denote two
plus two terms of the same form, but with different partitioning of
the rapidities, which can be obtained by cyclic permutation from those
displayed. We also used the exclusion property mentioned after eqn.
(\ref{eq:Fs_definition}).

We can now proceed by collecting terms according to the number of
free rapidity variables. The terms containing threefold summation
are\begin{eqnarray*}
 &  & \frac{1}{6}\sum_{123}\mathrm{e}^{-R(E_{1}+E_{2}+E_{3})}\left(\langle123|\mathcal{O}|123\rangle_{L}-\langle\mathcal{O}\rangle_{L}\right)-\frac{1}{2}\sum_{1}\sum_{2,3}\left(\langle23|\mathcal{O}|23\rangle_{L}-\langle\mathcal{O}\rangle_{L}\right)\\
 & + & \left(\sum_{1}\sum_{2}\sum_{3}-\frac{1}{2}\sum_{1,2}\sum_{3}\right)\left(\langle3|\mathcal{O}|3\rangle_{L}-\langle\mathcal{O}\rangle_{L}\right)\end{eqnarray*}
Replacing the sums with integrals\begin{eqnarray*}
\sum_{1} & \rightarrow & \int\frac{d\theta_{1}}{2\pi}\rho(1)\\
\sum_{1,2} & \rightarrow & \int\frac{d\theta_{1}}{2\pi}\frac{d\theta_{2}}{2\pi}\rho(12)\\
\sum_{1,2,3} & \rightarrow & \int\frac{d\theta_{1}}{2\pi}\frac{d\theta_{2}}{2\pi}\frac{d\theta_{3}}{2\pi}\rho(123)\end{eqnarray*}
and using (\ref{eq:matelms}) we get\begin{eqnarray*}
 &  & \frac{1}{6}\int\frac{d\theta_{1}}{2\pi}\frac{d\theta_{2}}{2\pi}\frac{d\theta_{3}}{2\pi}\mathrm{e}^{-R(E_{1}+E_{2}+E_{3})}\left(F(123)+3\rho(1)F(23)+3\rho(12)F(3)\right)\\
 & - & \frac{1}{2}\int\frac{d\theta_{1}}{2\pi}\frac{d\theta_{2}}{2\pi}\frac{d\theta_{3}}{2\pi}\mathrm{e}^{-R(E_{1}+E_{2}+E_{3})}\left(\rho(1)F(23)+2\rho(1)\rho(2)F(3)\right)\\
 & + & \int\frac{d\theta_{1}}{2\pi}\frac{d\theta_{2}}{2\pi}\frac{d\theta_{3}}{2\pi}\mathrm{e}^{-R(E_{1}+E_{2}+E_{3})}\left(\rho(1)\rho(2)F(3)-\frac{1}{2}\rho(12)F(3)\right)\end{eqnarray*}
where we reshuffled some of the integration variables. Note that all
terms cancel except the one containing $F(123)$ and writing it back
to its usual form we obtain\begin{equation}
\frac{1}{6}\int\frac{d\theta_{1}}{2\pi}\frac{d\theta_{2}}{2\pi}\frac{d\theta_{3}}{2\pi}F_{6}^{s}(\theta_{1},\theta_{2},\theta_{3})\mathrm{e}^{-mR(\cosh\theta_{1}+\cosh\theta_{2}+\cosh\theta_{3})}\label{eq:res3int}\end{equation}
It is also easy to deal with terms containing a single integral. The
only term of this form is\[
\frac{1}{3}\sum_{1=2=3}\mathrm{e}^{-R(E_{1}+E_{2}+E_{3})}\left(\langle123|\mathcal{O}|123\rangle_{L}-\langle\mathcal{O}\rangle_{L}\right)\]
When all rapidities $\theta_{1}^{(3)},\theta_{2}^{(3)},\theta_{3}^{(3)}$
are equal, the three-particle Bethe-Yang equations reduce to the one-particle
case\[
mL\sinh\theta_{1}^{(3)}=2\pi I_{1}\]
Therefore the relevant state density is that of the one-particle state:\begin{eqnarray}
\frac{1}{3}\int\frac{d\theta_{1}}{2\pi}\mathrm{e}^{-3RE_{1}}\rho(1)\left(\langle111|\mathcal{O}|111\rangle_{L}-\langle\mathcal{O}\rangle_{L}\right) & = & \int\frac{d\theta_{1}}{2\pi}\mathrm{e}^{-3RE_{1}}\rho(1)\frac{\rho(11)}{\rho(111)}F(1)\nonumber \\
 & \rightarrow & \int\frac{d\theta_{1}}{2\pi}\mathrm{e}^{-3mR\cosh\theta{}_{1}}F_{2}^{s}\label{eq:res1int}\end{eqnarray}
where we used that\[
\rho(1)\frac{\rho(11)}{\rho(111)}\rightarrow1\]
when $L\rightarrow\infty$.

The calculation of double integral terms is much more involved. We
need to consider\begin{eqnarray}
 &  & -\frac{1}{6}\left(\sum_{1=2,3}+\sum_{1=3,2}+\sum_{2=3,1}\right)\mathrm{e}^{-R(E_{1}+E_{2}+E_{3})}\left(\langle123|\mathcal{O}|123\rangle_{L}-\langle\mathcal{O}\rangle_{L}\right)\nonumber \\
 &  & +\frac{1}{2}\sum_{1}\sum_{2=3}\mathrm{e}^{-R(E_{1}+E_{2}+E_{3})}\left(\langle23|\mathcal{O}|23\rangle_{L}-\langle\mathcal{O}\rangle_{L}\right)\nonumber \\
 &  & +\frac{1}{2}\sum_{1=2}\sum_{3}\mathrm{e}^{-R(E_{1}+E_{2}+E_{3})}\left(\langle3|\mathcal{O}|3\rangle_{L}-\langle\mathcal{O}\rangle_{L}\right)\label{eq:dintstart}\end{eqnarray}
We need the density of partially degenerate three-particle states.
The relevant Bethe-Yang equations are\begin{eqnarray*}
mL\sinh\theta_{1}+\delta(\theta_{1}-\theta_{2}) & = & 2\pi I_{1}\\
mL\sinh\theta_{2}+2\delta(\theta_{2}-\theta_{1}) & = & 2\pi I_{2}\end{eqnarray*}
where we supposed that the first and the third particles are degenerate
(i.e. $I_{3}=I_{1}$), and used a convention for the phase-shift and
the quantum numbers where $\delta(0)=0$. The density of these degenerate
states is then given by\[
\bar{\rho}(13,2)=\det\left(\begin{array}{ll}
LE_{1}+\varphi(\theta_{1}-\theta_{2}) & -\varphi(\theta_{1}-\theta_{2})\\
-2\varphi(\theta_{1}-\theta_{2}) & LE_{2}+2\varphi(\theta_{1}-\theta_{2})\end{array}\right)\]
where we used that $\varphi(\theta)=\varphi(-\theta)$. Using the
above result and substituting integrals for the sums, we can rewrite
eqn. (\ref{eq:dintstart}) in the form\begin{eqnarray*}
 &  & -\frac{1}{6}\int\frac{d\theta_{1}}{2\pi}\frac{d\theta_{2}}{2\pi}\mathrm{e}^{-R(2E_{1}+E_{2})}\frac{\bar{\rho}(13,2)}{\rho(112)}\left(2\rho(1)F(12)+2\rho(12)F(1)+\rho(11)F(2)\right)+\dots\\
 &  & +\frac{1}{2}\int\frac{d\theta_{1}}{2\pi}\frac{d\theta_{2}}{2\pi}\mathrm{e}^{-R(E_{1}+2E_{2})}\rho(1)\rho(2)\frac{2\rho(2)}{\rho(22)}F(2)\\
 &  & +\frac{1}{2}\int\frac{d\theta_{1}}{2\pi}\frac{d\theta_{3}}{2\pi}\mathrm{e}^{-R(2E_{1}+E_{3})}\rho(1)\rho(3)\frac{1}{\rho(3)}F(3)\end{eqnarray*}
where the ellipsis denote two terms that can be obtained by cyclical
permutation of the indices $1,2,3$ from the one that is explicitly
displayed, and these three contributions can be shown to be equal
to each other by relabeling the integration variables: \begin{eqnarray}
 &  & -\frac{1}{2}\int\frac{d\theta_{1}}{2\pi}\frac{d\theta_{2}}{2\pi}\mathrm{e}^{-R(2E_{1}+E_{2})}\frac{\bar{\rho}(13,2)}{\rho(112)}\left(2\rho(1)F(12)+2\rho(12)F(1)+\rho(11)F(2)\right)\nonumber \\
 &  & +\frac{1}{2}\int\frac{d\theta_{1}}{2\pi}\frac{d\theta_{2}}{2\pi}\mathrm{e}^{-R(E_{1}+2E_{2})}\rho(1)\rho(2)\frac{2\rho(2)}{\rho(22)}F(2)\nonumber \\
 &  & +\frac{1}{2}\int\frac{d\theta_{1}}{2\pi}\frac{d\theta_{3}}{2\pi}\mathrm{e}^{-R(2E_{1}+E_{3})}\rho(1)\rho(3)\frac{1}{\rho(3)}F(3)\label{eq:dintreshuff}\end{eqnarray}
We first evaluate the terms containing $F(23)$ which results in\begin{equation}
-\int\frac{d\theta_{1}}{2\pi}\frac{d\theta_{2}}{2\pi}F_{4}^{s}(\theta_{1},\theta_{2})\mathrm{e}^{-mR(\cosh\theta_{1}+2\cosh\theta_{2})}\label{eq:dintres1}\end{equation}
using that \[
\frac{\bar{\rho}(13,2)}{\rho(112)}\rho(1)=1+O(L^{-1})\]
We can now treat the terms containing the amplitude $F(1)=F(2)=F(3)=F_{2}^{s}$.
Exchanging the variables $\theta_{1}\leftrightarrow\theta_{2}$ in
the second line and redefining $\theta_{3}\rightarrow\theta_{2}$
in the third line of eqn. (\ref{eq:dintreshuff}) results in\[
\frac{F_{2}^{s}}{2}\int\frac{d\theta_{1}}{2\pi}\frac{d\theta_{2}}{2\pi}\mathrm{e}^{-R(2E_{1}+E_{2})}\left\{ -\frac{\bar{\rho}(13,2)}{\rho(112)}\left(2\rho(12)+\rho(11)\right)+\frac{2\rho(1)^{2}\rho(2)}{\rho(11)}+\rho(1)\right\} \]
The combination of the various densities in this expression requires
special care. From the large $L$ asymptotics\[
\rho(i)\sim E_{i}L\quad,\quad\rho(ij)\sim E_{i}E_{j}L^{2}\quad,\quad\rho(ijk)\sim E_{i}E_{j}E_{k}L^{3}\quad,\quad\bar{\rho}(13,2)\sim E_{1}E_{2}L^{2}\]
it naively scales with $L$. However, it can be easily verified that
the coefficient of the leading term, which is linear in $L$, is exactly
zero. Without this, the large $L$ limit would not make sense, so
this is rather reassuring. We can then calculate the subleading term,
which requires tedious but elementary manipulations. The end result
turns out to be extremely simple\begin{equation}
-\frac{\bar{\rho}(13,2)}{\rho(112)}\left(2\rho(12)+\rho(11)\right)+\frac{2\rho(1)^{2}\rho(2)}{\rho(11)}+\rho(1)=-\varphi(\theta_{1}-\theta_{2})+O(L^{-1})\label{eq:anomdens}\end{equation}
so the contribution in the $L\rightarrow\infty$ limit turns out to
be just\begin{equation}
-\frac{1}{2}\int\frac{d\theta_{1}}{2\pi}\frac{d\theta_{2}}{2\pi}F_{2}^{s}\varphi(\theta_{1}-\theta_{2})\mathrm{e}^{-mR(2\cosh\theta_{1}+\cosh\theta_{2})}\label{eq:dintres2}\end{equation}
Summing up the contributions (\ref{eq:res3int}), (\ref{eq:res1int}),
(\ref{eq:dintres1}) and (\ref{eq:dintres2}) we indeed obtain (\ref{eq:our3order}).

\end{document}